\documentclass[%
prc,
aps,
showpacs,
showkeys,
] {revtex4}
\usepackage{graphicx,hyperref,amssymb,amsmath}
\usepackage{mathrsfs} 
\begin{document}

\title{Neutrino accompanied $\beta^{\pm}\beta^{\pm}$,
$\beta^+/EC$ and $EC/EC$ processes 
within single state dominance hypothesis}

\date{\today}

\author{P. Domin}
\email{pavol.domin@usm.cl}
\affiliation{Departamento de F{\' i}sica, Universidad T{\' e}cnica Federico 
Santa Mar{\' i}a, Casilla 110-V, Valpara{\' i}so, Chile}

\author{S. Kovalenko}
\email{sergey.kovalenko@usm.cl}
\affiliation{Departamento de F{\' i}sica, Universidad T{\' e}cnica Federico
Santa Mar{\' i}a, Casilla 110-V, Valpara{\' i}so, Chile}

\author{F. \v Simkovic}
\email{fedor.simkovic@fmph.uniba.sk}
\affiliation{Department of Nuclear Physics, Comenius University, Bratislava, 
Slovakia}

\author{S. V. Semenov}
\email{semenov@imp-server.imp.kiae.ru}
\affiliation{Russian Research Center ``Kurchatov Institute'', Moscow, Russia}

\begin{abstract}
The systematic study of (anti-)neutrino 
accompanied $\beta^{\pm}\beta^{\pm}$ decays
and $\beta^+/EC$, $EC/EC$ electron captures
is performed under the assumption
of single intermediate nuclear state dominance. The corresponding
half-lives are evaluated both for transitions 
to the ground state as well as to the $0^+$ and
$2^+$ excited states of final nucleus.  
It is stressed 
that the hypothesis of single state dominance can be confirmed 
or ruled out by the precision measurements of the 
differential characteristics of the
$2\nu\beta^-\beta^-$-decays of $^{100}Mo$ and 
$^{116}Cd$ as well as $\beta^+/EC$ electron capture in   
$^{106}Cd$, $^{130}Ba$ and $^{136}Ce$.
\end{abstract}



\maketitle

\section{\label{sec:intro}Introduction}

Double beta decay is of a major importance both for particle and nuclear
physics. 
In particular, the lepton number violating neutrinoless mode of 
double beta decay ($0\nu\beta\beta$-decay), 
forbidden in the Standard Model(SM), touches upon one of the key 
questions of particle physics: if massive neutrinos are 
Majorana or Dirac particles. As is known this process occurs if 
and only if neutrinos are Majorana particles. 
Therefore any experimental observation of 
$0\nu\beta\beta$-decay would be an unambiguous indication 
of the Majorana nature of neutrinos even if neutrino contribution to 
its rate is negligible in comparison with some other mechanisms 
related to physics beyond the SM \cite{sch82}.   
It is commonly believed that the future double beta decay experiments 
have realistic chances to observe $0\nu\beta\beta$-decay and, thus, to 
establish the Majorana nature of neutrinos, provide an important 
information on neutrino mass matrix and on the lepton number 
violating parameters of physics beyond the SM.
The accuracy of the determination of these quantities
from the measurements of the $0\nu\beta\beta$-decay half-life will be 
essentially related to our knowledge of the corresponding nuclear 
matrix elements. Therefore, the development of reliable models for 
calculation of these matrix elements becomes more and more challenging 
problem of nuclear physics. One of the guiding principles for
construction of such models consists in testing their predictions for 
the SM allowed two neutrino double beta decay 
($2\nu\beta\beta$-decay) \cite{fae98c,suh98a} which 
has been directly observed in ten nuclides \cite{data}, including decays 
into two excited states ($^{100}Mo$\cite{mo100e} and $^{150}Nd$ \cite{zetf}).

However, the calculation of the $2\nu\beta\beta$-decay matrix elements 
is a complex task as the nuclei of interest are medium and heavy open shell
ones. An additional complication arises from the fact that this is a second 
order process in weak interactions and, therefore, the construction of the 
complete set of states of intermediate nucleus is needed.
At present there are basically two different approaches to the calculation of 
the $\beta\beta$-decay matrix elements which are the Quasiparticle Random 
Phase Approximation (QRPA) and the shell model \cite{fae98c,suh98a}. 
Both approaches have their specific advantages and disadvantages.  
In particular, the shell model is restricted to configurations with 
low-lying excitations, but takes into account all the possible correlations.
On the other hand, the model space of the QRPA includes orbits far away from
the Fermi surface, what allows one to describe high-lying  excited states 
(up to 20 or 30 MeV) of the intermediate nucleus. However, it is questionable 
whether this approach takes into account the relevant ground state 
correlations in a proper way.
Also, the  $2\nu\beta\beta$-decay matrix element evaluated within the QRPA
exhibits strong sensitivity to the details of nuclear Hamiltonian
and dependence upon the model assumptions. 
The shell model has been found successful in the case of the
$2\nu\beta\beta$-decay of $^{48}Ca$ \cite{ca48}. However, for other 
nuclear systems the shell model predictions are not convincing 
enough. In particular, for heavier nuclei it is not clear whether 
the considered model space is sufficiently large. 
This question becomes irrelevant only in the case when the dominant 
contribution to the  $2\nu\beta\beta$-decay matrix element comes from
the transitions through the lowest states of intermediate nucleus.
Good candidates with this property are especially those nuclear systems 
for which the ground state of the intermediate nucleus is $1^+$ state, 
e.g., $^{100}Mo$, $^{116}Cd$ and $^{128}Te$. 

Some times ago it was suggested by Abad et al. ~\cite{aba84} that 
the $2\nu\beta\beta$-decays with $1^+$ ground state of the 
intermediate nucleus are solely determined
by the two virtual beta decay transitions: i) the first one connecting
the ground state of the initial nucleus with $1^+_1$ intermediate
state; ii) the second one proceeding from the $1^+_1$ state to the
final ground state. This assumption is known as the single state
dominance (SSD) hypothesis. 
The main advantage of the phenomenological approach based on the 
SSD hypothesis is that it is model independent since the values 
of the single $\beta$-decay matrix elements can be deduced from the measured 
$log~ft$ values. 

The available experimental data\cite{gar93, eji96, aki97,bha98}
and the results of theoretical analysis \cite{grif,civ98,sem00} 
support the conjecture that the SSD hypothesis is valid for 
two-neutrino accompanied double beta decay. 
Recently, it has been suggested that the
SSD hypothesis can be confirmed or ruled out by the precision 
measurement of differential rates in the case of
$2\nu\beta\beta$-decay of $^{100}Mo$ \cite{sim01}. The 
corresponding effects have been studied for the 
NEMO 3 experiment. 

Previous theoretical studies \cite{grif,civ98} of the SSD predictions 
were based on the approximation, when the sum of two lepton energies 
in the denominator of the $2\nu\beta\beta$-decay matrix element is replaced 
with their average value. In Ref. \cite{sem00} it was shown that this 
treatment can be inadequate for some nuclear systems and demonstrated 
how to calculate the matrix elements without this approximation.

In this article we present the results of systematic
study of the neutrino emitting modes of double beta decay within the 
SSD approach without any additional approximation. The calculated
decay rates are given both for transitions to the ground and excited 
final states. We also examine the question which of double beta 
decaying nuclei have the best prospects for experimental verification 
of the SSD hypothesis via the analysis of the differential rates.  

In Section \ref{sec:theory} we describe our SSD based approach. The double 
beta decay rates are derived in Section \ref{sec:rates} Results and 
discussions are presented in Section \ref{sec:res}. Conclusions are drawn 
in Section \ref{sec:sum}.

\section{\label{sec:theory} SSD approach to double beta decay}

We distinguish four types of neutrino (antineutrino) accompanied nuclear 
double beta decays ($\beta\beta$-decays). The most attention has been 
previously paid to the mode with emission of two electrons 
($2\nu\beta^{-}\beta^{-}$-decay), 
\begin{equation}
(Z-2,A) \rightarrow (Z,A) + 2e^-  + 2{\tilde{\nu}}_{e},
\end{equation}
where Z and A denote atomic and mass numbers of a daughter nucleus,
respectively. Another possible mode is double beta decay
with emission of two positrons 
($2\nu\beta^{+}\beta^{+}$-decay),
\begin{equation}
(Z+2,A) \rightarrow (Z,A) + 2e^+ + 2{{\nu}}_{e}. 
\end{equation}
In comparison with the $2\nu\beta^{-}\beta^{-}$-decay 
this process is disfavored by smaller available kinetic
energy and also by the Coulomb repulsion on the positron. 
More favorable for detection than the  
$2\nu\beta^{+}\beta^{+}$-decay are the
capture processes, namely
the capture of one bound atomic electron with emission
of positron 
($2\nu\varepsilon\beta^{+}$-decay)     
\begin{equation}
e^-_b + (Z+2,A) \rightarrow (Z,A) + e^+ + 2{\nu}_{e}, 
\end{equation}
and the double capture of two bound atomic electrons 
($2\nu\varepsilon\varepsilon$-decay)
\begin{equation}
2 e^-_b + (Z+2,A) \rightarrow (Z,A) + 2 {{\nu}}_{e}. 
\end{equation}
The enhancement of the capture processes in comparison with the
$2\nu\beta^{+}\beta^{+}$-decay originates from the larger available 
energies and from the fact that the suppression
due to the Coulomb repulsion is partially 
($2\nu\varepsilon\beta^{+}$-decay) or fully
($2\varepsilon\varepsilon$-decay) avoided. 
The electron capture from the s-states is a dominant subprocess, 
because the corresponding electron wavefunction does not vanish at 
the origin. In practice, one takes into account the capture both from
the $K$ and $L_I$ shells. However, the probability of capture from $L_I$ 
shell is suppressed by about a factor 10 in comparison
with capture from $K$ shell. The inclusion of the electron capture processes 
into our analysis is, in particular, justified by a reviving 
interest to the experimental study of these processes \cite{TGV}.

The nuclear double beta decay is a second-order process of 
perturbation theory in weak interaction.
The standard $\beta$-decay Hamiltonian has the form
\begin{equation}
  \mathscr{H}^{\beta}(x)= - \frac{G_{\beta}}{\sqrt{2}} \bar e(x)
  \gamma^{\mu} (1 - \gamma_5) \nu_e(x) j_{\mu}(x) + \mathrm{h.c.},
  \label{eq:Hbeta}
\end{equation}
where $G_{\beta} = G_F \cos \theta_c$ and $G_F$ is Fermi constant,
$\theta_c$ is Cabbibo angle, $e(x)$, $\nu(x)$ are respectively electron and
neutrino field operators, $j_{\mu}(x)$ is strangeness-conserving free
charged hadron current. 

The $\beta\beta$-decay matrix elements involve summation over
the discrete states and integration over the continuum states of
the intermediate nucleus
in the presence of energy denominators corresponding to the second order of 
perturbation theory in weak interactions. 
The neutrino accompanied modes of $\beta\beta$-decay are dominated by 
the Gamow-Teller transitions allowed by the angular-momentum, parity and 
isospin selection rules. The contribution of double Fermi transition to 
the double beta decay amplitude has been found to be small \cite{hax84}.  
As is known, in nuclear models, like shell model and QRPA, the evaluation 
of the  nuclear matrix elements is restricted to the two subsequent 
Gamow-Teller transitions through a certain number of low-lying discrete 
states of the intermediate nucleus.

The partial $\beta\beta$-decay matrix elements associated with the transition
to the ground or excited states of the final nucleus with 
angular momentum and parity $J^\pi=0^+, 2^+$
can be written as 
\begin{equation}
\frac{1}{\sqrt{s}}
\sum_{m}
\frac{
  \langle J^{\pi}_f \parallel
  \sum_{{j}}\tau^{{\mp}}_{{j}}{\sigma}_{{j}} \parallel 1^+_{m} 
  \rangle
  \langle 1^+_{m} \parallel 
  \sum_{{j'}}\tau^{{\mp}}_{{j'}}{\sigma}_{{j'}} \parallel 0^+_i\rangle}
{\varepsilon_{{m}}-\varepsilon_{{i}}+ \varepsilon_j + \nu_k} ~~(j,k~=~1,2)
\label{mele}
\end{equation}
with $s=1$ for $J^\pi = 0^+$ and $s=3$ for $J^\pi = 2^+$. 
%
$|0^+_i\rangle$, $|J^{\pi}_f\rangle$ and $|1^+_{m} \rangle$ 
are respectively the wave functions of parent, daughter
and intermediate nuclei with
the corresponding energies $\varepsilon_{{i}}$, $\varepsilon_{{f}}$ 
and $\varepsilon_{{m}}$. The terms $\varepsilon_j$  ($j=1,2$) represent 
the energy of outgoing electron/positron 
or (with the minus sign) the energy of bound electron and $\nu_k$ ($k=1,2$)
is the energy of emitted neutrino or antineutrino. Both, nuclear and
lepton energies are assumed to be in the units of electron mass $m_e$.

For the sake of simplicity the sum of outgoing lepton energies
in the denominators of the $\beta\beta$ matrix elements 
is usually replaced with
its average value $f w_0$, where $f=1/2$ for $2\nu\beta^{\pm}\beta^{\pm}$ and
$2\nu\varepsilon\varepsilon$-modes and
$f=1/3$ or $f=2/3$ for $2\nu\varepsilon\beta^{+}$-mode 
($w_0 = (\varepsilon_{{i}}$ - $\varepsilon_{{f}})= W_0/m_e$,
see Table \ref{tab:bbdata}).
This approximation allows one to factorize the lepton and nuclear
parts in the calculation of the $\beta\beta$-decay rate.
It works generally quite good for the cases when
the dominant contribution to the $\beta\beta$ matrix elements comes 
from the transitions through the higher lying states of the intermediate 
nucleus.

The SSD hypothesis drastically 
simplifies the calculation of the two-neutrino $\beta\beta$ nuclear
matrix elements. It is supposed that nuclear matrix element
is governed by the two virtual transitions: the first one going from 
the initial ground state to the $1^+$  ground state of the intermediate 
nucleus and second one from this $1^+$ state to the final 
$J^\pi=0^+$ or $2^+$ state. 
With this assumption the relevant nuclear matrix element is as
follows:
\begin{equation}
  M^{(\pm)}_{GT,m=g.s.}(J^{\pi}) = \frac{1}{\sqrt{s}}
  \langle J^{\pi}_f \parallel
  \sum_{{j}}\tau^{{\mp}}_{{j}}{\sigma}_{{j}} \parallel 1^+_{g.s.} 
  \rangle
  \langle 1^+_{g.s.} \parallel 
  \sum_{{j'}}\tau^{{\mp}}_{{j'}}{\sigma}_{{j'}} \parallel 0^+_i\rangle.
\label{eq:MEL}
\end{equation}
The superscript $-$ ($+$) indicates the neutrino (antineutrino) 
emitting mode of the $\beta\beta$ decay. 
The value of this matrix element can be determined in a model independent way 
from the single $\beta$ decay and electron capture measurements.
>From the experimental values of $\log ft$ for electron capture 
and single $\beta$ decay of the  ground state of the intermediate 
nucleus with $J^\pi=1^+$ we get
\begin{subequations}
  \label{eq:SSDMEL}
 \begin{align}
  &\langle J^{\pi}_f \parallel
  \sum_{{j}}\tau^{{-}}_{{j}}{\sigma}_{{j}} \parallel 1^+_{\mathrm{g.s.}}
  \rangle =
  \frac{1}{g_A}\sqrt{\frac{3D}{ft_{\beta^-}}},& 
  &\langle 1^+_{\mathrm{g.s.}} \parallel 
  \sum_{{j'}}\tau^{{-}}_{{j'}}{\sigma}_{{j'}} \parallel 0^+_i\rangle=
  \frac{1}{g_A}\sqrt{\frac{3D}{ft_{\varepsilon}}},\\
  &\langle J^{\pi}_f \parallel
  \sum_{{j}}\tau^{{+}}_{{j}}{\sigma}_{{j}} \parallel 1^+_{\mathrm{g.s.}} 
  \rangle=
  \frac{1}{g_A}\sqrt{\frac{3D}{ft_{\varepsilon}}},&
  &\langle 1^+_{\mathrm{g.s.}} \parallel 
  \sum_{{j'}}\tau^{{+}}_{{j'}}{\sigma}_{{j'}} \parallel 0^+_i\rangle=
  \frac{1}{g_A}\sqrt{\frac{3D}{ft_{\beta^{-}}}},
 \end{align}
\end{subequations}
where $D=(3 \pi^3 \ln 2)/(G_{\beta}^2 m_e^5)$ is beta decay constant.
The $\beta\beta$-emitters, which fulfil the SSD hypothesis
condition, namely those with $J^\pi = 1^+$  
ground state of the intermediate nucleus,
are listed in Tables \ref{tab:bbdata} and \ref{tab:eedata}.

\begin{table}[!ht]
  \caption{Basic characteristics of $\beta^-\beta^-$ unstable nuclei,
with $1^+$ spin-parity of the ground state of intermediate nucleus.
$W_0 = E_i - E_f$ is the energy difference between initial and final 
ground states in MeV
deduced from atomic masses ~\protect\cite{aud95}.
$\Delta = E_1 - E_i~$ with $E_1$ being the energy of the
ground state of intermediate nucleus. 
$\delta$ is the natural abundance of the isotope. 
  \label{tab:bbdata}} 
  \begin{center}
\begin{tabular}{lllllllll}
  \hline
  \hline
&
${^{70}\mathrm{Zn}}$
&${^{80}\mathrm{Se}}$
&${^{100}\mathrm{Mo}}$
&${^{104}\mathrm{Ru}}$
&${^{110}\mathrm{Pd}}$
&${^{114}\mathrm{Cd}}$
&${^{116}\mathrm{Cd}}$
&${^{128}\mathrm{Te}}$\\ 
 \hline
$W_0$ [MeV]&
$2.0229$&
$1.1559$&
$4.0563$&
$2.3216$&
$3.0217$&
$1.5588$&
$3.8270$&
$1.8892$\\ 
$\Delta$ [MeV]&
$0.1437$&
$1.3596$&
$-0.3429$&
$0.6302$&
$0.3814$&
$0.9409$&
$-0.0410$&
$0.7408$\\ 
$\delta$&
$0.62$&
$49.61$&
$9.63$&
$18.62$&
$11.72$&
$28.73$&
$17.49$&
$31.74$\\
\hline
\hline
\end{tabular}
\end{center}
\end{table}

\begin{table}[!ht]
  \caption{
Basic characteristics of experimentally interesting double electron 
capture unstable nuclei 
with $1^+$ spin-parity of the ground state of intermediate nucleus.
$E_e(K)$ and  $E_e(L)$  are the energies of bound 
electron for $K$ ($1s_{1/2}$) and $L_{I}$ ($2s_{1/2})$ shells, respectively.
$\mathfrak{N}_{0,-1}$ ($\mathfrak{N}_{0,-1}$) denotes the probability to find 
$1s_{1/2}$ ($2s_{1/2}$) state electron 
in the volume $1/m^3_e$ around the origin of nucleus 
(see Eqs.~\eqref{eq:FacNa} and   \eqref{eq:FacNb}).
$W_0$, $\Delta$, $E_e(K)$ and  $E_e(L)$  are given in units of MeV.
The quantities $W_0$, $\Delta$, $\delta$ are defined in Table 
\protect\ref{tab:bbdata}.
  \label{tab:eedata}} 
  \begin{center}
    \begin{tabular}{lllllllllll}
      \hline
      \hline
&
${^{64}\mathrm{Zn}}$&
${^{78}\mathrm{Kr}}$&
${^{106}\mathrm{Cd}}$&
${^{108}\mathrm{Cd}}$&
${^{112}\mathrm{Sn}}$&
${^{120}\mathrm{Te}}$&
${^{130}\mathrm{Ba}}$&
${^{136}\mathrm{Ce}}$&
${^{162}\mathrm{Er}}$&
 ${^{164}\mathrm{Er}}$\\ 
      \hline
$W_0$&
$0.0743$&
$1.8440$&
$1.7491$&
$-0.7528$&
$0.9002$&
$0.6764$&
$1.5886$&
$1.3752$&
$0.8225$&
$-0.9979$\\ 
$\Delta$&
$1.0897$&
$1.2188$&
$0.7051$&
$2.1600$&
$1.1747$&
$1.4932$&
$0.8796$&
$0.9838$&
$0.8067$&
$1.4738$\\ 
$\delta$&
$48.63$&
$0.35$&
$1.25$&
$0.89$&
$0.97$&
$0.09$&
$0.106$&
$0.185$&
$0.14$&
$1.61$\\
$E_e(K)$&
$0.4986$&
$0.4931$&
$0.4786$&
$0.4786$&
$0.4758$&
$0.4728$&
$0.4664$&
$0.4630$&
$0.4437$&
$0.4437$\\ 
$E_e(L)$&
$0.5079$&
$0.5065$&
$0.5028$&
$0.5028$&
$0.5021$&
$0.5014$&
$0.4997$&
$0.4988$&
$0.4939$&
$0.4939$\\ 
$\mathfrak{N}_{0,-1}$&
$4.521~10^{-3}$&
$8.792~10^{-3}$&
$2.776~10^{-2}$&
$2.774~10^{-2}$&
$3.311~10^{-2}$&
$3.934~10^{-2}$&
$5.524~10^{-2}$&
$6.522~10^{-2}$&
$1.467~10^{-1}$&
$1.465~10^{-1}$\\ 
$\mathfrak{N}_{1,-1}$&
$5.896~10^{-4}$&
$1.168~10^{-3}$&
$3.872~10^{-3}$&
$3.869~10^{-3}$&
$4.662~10^{-3}$&
$5.592~10^{-3}$&
$8.020~10^{-3}$&
$9.574~10^{-3}$&
$2.291~10^{-2}$&
$2.288~10^{-2}$\\
\hline
\hline
\end{tabular}
\end{center}
\end{table}

In the previous SSD studies of neutrino (antineutrino) $\beta\beta$-decay the
approximate treatment of the energy denominators has been applied 
\cite{grif,civ98}.
The sum of lepton energies entering the energy denominators were
replaced with a half of the energy release of this process.
Recently, it was shown that this approximation may lead to a 
significant inaccuracy in determining $\beta\beta$-decay
half-lives for the transition to ground and  excited states
\cite{sim01}. 
It was demonstrated that the exact treatment of the energy denominators 
is especially important 
for $\beta\beta$-nuclear systems with a small energy difference between 
the intermediate and initial ground 
states in comparison with the value of $W_0$, e.g., A=100 and 116
systems.  

The effect of energy denominators may have a strong impact on the 
differential rates. In Ref. \cite{sim01} the two limiting cases 
have been considered:
i) The differential rates have been evaluated within the SSD hypothesis
with the exact treatment of the energy denominators, which are the functions
of lepton energies. 
ii) The above mentioned standard approximation for the energy 
denominators has been
applied in which case the energy denominators are independent 
of lepton energies.

As we commented above, this approximation is justified if the 
transitions through the higher lying states of the intermediate 
nucleus give the dominant contribution to the 
$\beta\beta$-decay amplitude. This assumption, which we call 
as the higher state dominance (HSD) hypothesis, is an alternative 
to the SSD hypothesis leading to the different
predictions for the differential rates. 
Thus, from the comparison of experimental data on differential rates with 
SSD and HSD predictions one can confirm or rule out the SSD hypothesis. 
This would provide us with a valuable information about the structure 
of involved nuclear matrix elements.

\section{\label{sec:rates}Double beta decay rates}

In what follows we present basic formulas needed for evaluation of the
$\beta\beta$-decay half-lives and the differential rates within the SSD 
and HSD hypotheses.

\subsection{The $2\nu\beta^{\pm}\beta^{\pm}$-decay half-life}

The differential rate for $2\nu\beta^-\beta^-$- and
$2\nu\beta^+\beta^+$-decays leading to $J^{\pi}$ 
($J^{\pi}$ = $0^{+}$, $2^{+}$) 
state of the daughter nucleus takes the form~\cite{doi85, doi92}
\begin{equation}
  d \omega^{2\nu\beta^{\pm}\beta^{\pm}}_{J^{\pi}} = a_{2\nu}
  {\cal A}^{2\nu\beta^{\pm}\beta^{\pm}}_{J^{\pi}}
(\varepsilon_1, \varepsilon_2, \nu_1, \nu_2)
  d \Omega,
  \label{eq:2vbbwidtha}
\end{equation}
where $a_{2\nu} = (G_{\beta} g_A)^4 m_e^9 / (64 \pi^7)$. 
The phase-space factor $d\Omega$ is 
\begin{equation}
  d\Omega = 2  \pi_1 \pi_2 \varepsilon_1 \varepsilon_2 \nu_1^2 \nu_2^2 
  \delta(\varepsilon_1 + \varepsilon_2 + \nu_1 + \nu_2 - \varepsilon_i +
  \varepsilon_f) 
  d \varepsilon_1 d \varepsilon_2 d \nu_1 d \nu_2.
\end{equation}
Here,  $\pi_k$ is the  momentum of k-th emitted beta particle. 
The electron/positron  momentum $\pi_k$, energy $\varepsilon_k$ and
neutrino energy $\nu_k$ ($k=1,2$) are assumed to be 
in the units of electron mass $m_e$. The function 
${\cal A}^{I}_{J^{\pi}}$ can be written as
\begin{equation}
  {\cal A}^{2\nu\beta^{\pm}\beta^{\pm}}_{J^{\pi}}
(\varepsilon_1, \varepsilon_2, \nu_1, \nu_2) =
  a(\varepsilon_1, \varepsilon_2)
  {\cal M}^{2\nu\beta^{\pm}\beta^{\pm}}_{J^{\pi}}
(\varepsilon_1, \varepsilon_2, \nu_1, \nu_2)
  \label{eq:FacA}
\end{equation}
with 
\begin{equation}
    a(\varepsilon_1, \varepsilon_2)=
    F^{(\pm)}_0(\varepsilon_1) R^{(\pm)}_{11}(\varepsilon_1)
    F^{(\pm)}_0(\varepsilon_2) R^{(\pm)}_{11}(\varepsilon_2),
\label{ecoulf}
\end{equation}
Here, the superscript $-$ ($+$) denotes 
$2\nu\beta^-\beta^-$($2\nu\beta^+\beta^+$) - decay mode. 
$F^{(-)}_0(\varepsilon)$ is the relativistic Coulomb 
factor~\cite{doi85}. The correction factor $R^{(-)}_{11}(\varepsilon)$ is
given by 
\begin{equation}
  \label{eq:R11}
  R^{(-)}_{11}(\varepsilon)=
  R_{-1}(\varepsilon) + R_{+1}(\varepsilon)
\end{equation}
with $R_{\pm k}(\varepsilon) = D^2_{\pm k}(\varepsilon) (\varepsilon \mp
m_e)/(2 \varepsilon)$. 
The function  $D^2_{\pm k}(\varepsilon)$ reflects the charge
distribution inside the nucleus. Its explicit form can be found in
Ref.~\cite{doi88b}. The calculation of $F^{(+)}_0(\varepsilon)$ and 
$R^{(+)}_{11}(\varepsilon)$ differs from the calculation of 
$F^{(-)}_0(\varepsilon)$  and  $R^{(-)}_{11}(\varepsilon)$ 
in the sign of Z associated with  the  final nucleus. 

The function ${\cal M}^{2\nu\beta^{\pm}\beta^{\pm}}_{J^{\pi}}
(\varepsilon_1, \varepsilon_2, \nu_1, \nu_2)$ in 
Eq. (\ref{eq:FacA}) is the product of the lepton energy dependent 
nuclear matrix elements.
Within the SSD hypothesis it takes the form 
\begin{equation}
  \label{eq:facMSSDa}
    {\cal M}^{2\nu\beta^{\pm}\beta^{\pm}}_{J^{\pi}}
(\varepsilon_1, \varepsilon_2, \nu_1, \nu_2) =
    |M^{(\pm)}_{GT, \mathrm{g.s.}}(J^{\pi})|^2 
    {\cal K}_{J^{\pi}}(\varepsilon_1, \varepsilon_2, \nu_1, \nu_2).
\end{equation}
The  nuclear matrix element 
$M^{(\pm)}_{GT, \mathrm{g.s.}}(J^{\pi})$ consists of
two single $\beta$ decay matrix elements associated with transitions
via $1^+$ ground state of the intermediate nucleus [see  Eq. (\ref{eq:MEL})].
The energy denominators enter into the expression for 
\begin{eqnarray}
  \label{eq:calK}
  {\cal K}_{J^{\pi}} (\varepsilon_1, \varepsilon_2, \nu_1, \nu_2)
&=&  \frac{1}{3}
  (K^2_{\mathrm{g.s.}}(\varepsilon_1, \varepsilon_2, \nu_1, \nu_2) +
  L^2_{\mathrm{g.s.}}(\varepsilon_1, \varepsilon_2, \nu_1, \nu_2) 
\nonumber\\
&&+ K_{\mathrm{g.s.}}(\varepsilon_1, \varepsilon_2, \nu_1, \nu_2) 
  L_{\mathrm{g.s.}}(\varepsilon_1, \varepsilon_2, \nu_1, \nu_2)) 
~~~ \ \ \ \ (for~~J^\pi=0^+),
\nonumber\\
&=&
  (K_{\mathrm{g.s.}}(\varepsilon_1, \varepsilon_2, \nu_1, \nu_2) - 
  L_{\mathrm{g.s.}}(\varepsilon_1, \varepsilon_2, \nu_1, \nu_2))^2
~~~(for~~J^\pi=2^+)
\end{eqnarray}
with 
\begin{eqnarray}
  K_{m=g.s.}(\varepsilon_1, \varepsilon_2, \nu_1, \nu_2)&=& 
  (\varepsilon_{m=g.s.} - \varepsilon_i + \varepsilon_1 + \nu_1)^{-1} +
  (\varepsilon_{m=g.s.} - \varepsilon_i + \varepsilon_2 + \nu_2)^{-1},\\
  L_{m=g.s.}(\varepsilon_1, \varepsilon_2, \nu_1, \nu_2)&=& 
  (\varepsilon_{m=g.s.} - \varepsilon_i + \varepsilon_1 + \nu_2)^{-1} +
  (\varepsilon_{m=g.s.} - \varepsilon_i + \varepsilon_2 + \nu_1)^{-1}.
\label{eq:denoms}
\end{eqnarray}
Here, $\varepsilon_{m=g.s.}$ corresponds to the energy of the $1^+$ ground 
state of the intermediate nucleus.

Let us note that the general matrix element used in the nuclear model dependent
calculation of the $2\nu\beta^\pm\beta^\pm$-decay rates includes the
sum over the intermediate nuclear states. 
In this case it is evaluated within the
standard approximation,  when 
in the common denominator of $K_m$ and $L_m$ 
the lepton energies are replaced with their average values 
\begin{equation}
  \varepsilon_1 \simeq 
  \varepsilon_2 \simeq 
  \nu_1 \simeq
  \nu_2 \simeq \frac{w_0^{(2\nu\beta^{\pm}\beta^{\pm})}}{4}, ~~~
    w_0^{(2\nu\beta^{\pm}\beta^{\pm})} = \varepsilon_i - \varepsilon_f,
\label{aver1}
\end{equation}
for the transitions to the $0^{+}$ final states. For the transitions to the
$2^{+}$ final states the above replacement is applied to the lepton energies 
in the denominator of the expression
\begin{equation}
  K_m - L_m = 
  \frac{2 (\nu_1 - \nu_2) (\varepsilon_1 - \varepsilon_2) 
  (\varepsilon_m - \varepsilon_i + (\varepsilon_i - \varepsilon_f)/2)}%
  {(\varepsilon_m - \varepsilon_i + \varepsilon_1 + \nu_1)
   (\varepsilon_m - \varepsilon_i + \varepsilon_2 + \nu_2)
   (\varepsilon_m - \varepsilon_i + \varepsilon_1 + \nu_2)
   (\varepsilon_m - \varepsilon_i + \varepsilon_2 + \nu_1)}.
\end{equation}
As the result of these approximations
the expression for ${\cal M}^{2\nu\beta^{\pm}\beta^{\pm}}_{J^{\pi}}$ 
in Eq. (\ref{eq:facMSSDa}) can be factorized into the product of 
the lepton-energy independent sum over the intermediate nuclear states and 
the factor that does not depends on $\varepsilon_m$.
Its explicit form for $J^\pi=0^+,~2^+$  is given in Table \ref{tab:FacHSD}.

Within the SSD hypothesis the total $2\nu\beta^{\pm}\beta^{\pm}$-decay rate
can be written as
\begin{equation}
\omega^{2\nu\beta^{\pm}\beta^{\pm}}_{J^\pi} = 
\frac{ln 2}{
T^{{2\nu\beta^{\pm}\beta^{\pm}}}_{\mathrm{SSD}}(J^{\pi})}
= c_{2\nu}^{SSD}
  \int\limits_{1}^{w_0^{(\alpha)} - 1} d \varepsilon_1
  \varepsilon_1 \pi_1
  \int\limits_{1}^{w_0^{(\alpha)} - \varepsilon_1} 
  d \varepsilon_2 \varepsilon_2 \pi_2
  \int\limits_{0}^{w_0^{(\alpha)} - \varepsilon_1 -
  \varepsilon_2} 
  d \nu_1 \nu_1^2 \nu_2^2 a(\varepsilon_1,  \varepsilon_2) 
  {\cal K}_{J^{\pi}}(\varepsilon_1, \varepsilon_2, \nu_1, \nu_2),
\label{omega+-}
\end{equation}
where $\alpha={{2\nu\beta^{\pm}\beta^{\pm}}}$.
The factor $c^{SSD}_{2\nu}$ determined from Eqs. (\ref{eq:MEL})
and (\ref{eq:SSDMEL}) takes the form
\begin{equation}
c^{SSD}_{2\nu}={2 a_{2\nu} |M^{\pm}_{GT,\mathrm{g.s.}}}|^2 
= \frac{9}{8\pi} \frac{(ln 2)^2}{m_e ft_\varepsilon~ft_{\beta^-}}.
\label{comega+-}
\end{equation}
It is worthwhile to notice that the 
$2\nu\beta^{\pm}\beta^{\pm}$-decay half-life in Eqs. (\ref{omega+-}) and
(\ref{comega+-}) does not depend explicitly on the values of $G_\beta$ and the
axial weak coupling constant $g_A$.

\begin{table}
\caption{%
General nuclear matrix elements
${\cal M}^{\alpha}_{J^{\pi}}$
entering into the expressions for $\alpha = 2\nu\beta^\pm\beta^\pm,
{2\nu\varepsilon\beta^{+}}, {2\nu\varepsilon\varepsilon}$-decay 
rates [see  Eqs. (\protect\ref{eq:FacA}),  (\protect\ref{eq:FaccA})
and   (\protect\ref{eq:2vbbwidthc})] with the standard approximation
in their common denominators. For the explanation and definition 
of $w_0^{(\alpha)}$ see Eqs. (\protect\ref{aver1}), 
(\protect\ref{aver2}) and (\protect\ref{aver3}). 
The quantities $\Delta_m = \varepsilon_m - \varepsilon_i$ include the energy 
$\varepsilon_m$ of intermediate $1^+$ nuclear state  and the 
ground state energy $\varepsilon_i$ of initial nucleus. 
The summation $\sum_m$ runs over the intermediate nuclear states. 
$\varepsilon_{e_l}$ with $l=1, 2$ stands for the energy of captured
electron.
$\varepsilon_l$ and  $\nu_k$ with $l, k=1, 2$ are the energies of
emitted  electron/positron and (anti)neutrino, respectively.
All the energies are in the units of $m_e$.
\label{tab:FacHSD}}
\begin{center}
\begin{tabular}{lll}
\hline
\hline\\
mode&
$J^{\pi}=0^+$&
$J^{\pi}=2^+$\\
\hline \\
$2\nu\beta^{\pm}\beta^{\pm}$&
$\left| \sum_m 
 M_{GT,m}^{(\pm)}(0^{+})
 \frac{2}
 {\Delta_m + 
 \frac{w^{(\alpha)}_0}{2}} \right|^2$&
 $\left| \sum_m M_{GT,m}^{(\pm)}(2^{+})
 \frac{2 (\varepsilon_1 - \varepsilon_2) (\nu_1 - \nu_2)}
 {\left( \Delta_m + \frac{w^{(\alpha)}_0}{2}\right)^3} \right|^2$
 \\
$2\nu\varepsilon\beta^{+}$ & 
$\left| \sum_m M_{GT,m}^{(+)}(0^{+})
 \left[
 \frac{1}
 {\Delta_{m} - \varepsilon_{e_1}
 + \frac{w^{(\alpha)}_0}{3}} 
 + \frac{1}
 {\Delta_{m} 
 + \frac{2 w^{(\alpha)}_0}{3}} \right]
 \right|^2$&
$\left| \sum_m M_{GT,m}^{(+)}(2^{+})
 \frac{2 (\varepsilon_1 - \varepsilon_2) (\nu_1 - \nu_2)
 (\Delta_m + (\varepsilon_i - \varepsilon_f)/2)}
 {\left( \Delta_m - \varepsilon_{e_1} + \frac{w^{(\alpha)}_0}{3}\right)^2
 \left( \Delta_m + \frac{2 w^{(\alpha)}_0}{3}\right)^2} \right|^2$ 
 \\
$2\nu \varepsilon\varepsilon$&
$\left| \sum_m M_{GT,m}^{(+)}(0^{+})
 \left[
 \frac{1}
 {\Delta_{m} - \varepsilon_{e_1}
 + \frac{w^{(\alpha)}_0}{2}} 
 + \frac{1}
 {\Delta_{m} - \varepsilon_{e_2} 
 + \frac{w^{(\alpha)}_0}{2}} \right]
 \right|^2$&
$\left| \sum_m M_{GT,m}^{(+)}(2^{+})
 \frac{2 (\varepsilon_{e_2} - \varepsilon_{e_1}) (\nu_1 - \nu_2)
 (\Delta_m + (\varepsilon_i - \varepsilon_f)/2)}
 {\left( \Delta_m - \varepsilon_{e_1} + \frac{w^{(\alpha)}_0}{2}\right)^2
  \left( \Delta_m - \varepsilon_{e_2} + \frac{w^{(\alpha)}_0}{2}\right)^2}
  \right|^2$\\ 
\hline
\hline
\end{tabular}
\end{center}
\end{table}

\subsection{ The $2\nu\varepsilon\beta^{+}$-decay half-life}

In the similar fashion we treat  
$2\nu\varepsilon\beta^{+}$-decay mode with $0^+$ and $2^+$ states
of final nucleus. In this case the differential decay rate can 
be written in the form~\cite{doi92}
\begin{equation}
  d \omega^{2\nu\varepsilon\beta^{+}}_{J^{\pi}} = a_{2\nu} 
  N^{2\nu\varepsilon\beta^{+}}
  {\cal A}^{2\nu\varepsilon\beta^{+}}_{J^{\pi}}
(-\varepsilon_{e_1}, \varepsilon_1, \nu_1, \nu_2)
  d \Omega.
  \label{eq:2vbbwidthb}
\end{equation}
The factor $N^{2\nu\varepsilon\beta^{+}}$ 
is associated with the normalization
of  wavefunction of bound electron, namely
\begin{equation}
  \label{eq:FacNa}
    N^{2\nu\varepsilon\beta^{+}} = 
    \left\{ 
    \begin{array}{ll}
      2 (2 \pi^2) \mathfrak{N}_{0,-1}& \text{$K(1s_{1/2})$ capture}\\ 
      2 (2 \pi^2) \mathfrak{N}_{1,-1}& \text{$L_{I}(2s_{1/2})$ capture}
    \end{array}
    \right.
\end{equation}
Here, $\mathfrak{N}_{0,-1}$ ($\mathfrak{N}_{1,-1}$) is the probability to find 
$1s_{1/2}$ ($2s_{1/2}$) state electron in the volume $1/m^3_e$ around 
the origin of nucleus. 
The details of the calculation of these factors can be found in 
~\cite{doi92}. In Table \ref{tab:eedata} we present their numerical 
values for some nuclei. One may note that for $L_I$-shell 
the above probability is suppressed by about a factor of $1/8$ in 
comparison with that for $K$-shell~\cite{doi92}. 
The $2\nu\varepsilon\beta^{+}$-phase-space
factor  is given by 
\begin{equation}
  d\Omega = 
  2 \pi_1 \varepsilon_1 \nu_1^2 \nu_2^2 
  \delta(\varepsilon_1 - \varepsilon_{e_1} + \nu_1 + \nu_2 - \varepsilon_i +
  \varepsilon_f) 
  d \varepsilon_1 d \nu_1 d \nu_2,
\end{equation}
where $\varepsilon_{e_1}$ is the energy brought to nucleus by 
a captured electron. Next, we have 
\begin{equation}
  {\cal A}^{2\nu\varepsilon\beta^{+}}_{J^{\pi}}
(-\varepsilon_{e_1}, \varepsilon_1, \nu_1, \nu_2)
 = a(\varepsilon_1) 
  {\cal M}^{2\nu\varepsilon\beta^{+}}_{J^{\pi}}
(-\varepsilon_{e_1}, \varepsilon_1, \nu_1, \nu_2)
  \label{eq:FaccA}
\end{equation}
with
\begin{equation} \label{eq:26}
  a(\varepsilon_1) = F_0^{(+)}(\varepsilon_1) R_{11}^{(+)}(\varepsilon_1).
\end{equation}
The SSD form of the matrix element is
\begin{equation}
  \label{eq:facMSSDb}
    {\cal M}^{2\nu\varepsilon\beta^{+}}_{J^{\pi}}
(-\varepsilon_{e_1}, \varepsilon_1, \nu_1, \nu_2)=
    |M^{(+)}_{GT, \mathrm{g.s.}}(J^{\pi})|^2 
    {\cal K}_{J^{\pi}}
    (\varepsilon_1, -\varepsilon_{e_1}, \nu_1, \nu_2),
\end{equation}
where the nuclear matrix element and the energy denominator in 
the r.h.s. of this equation are defined in Eqs. (\ref{eq:MEL}) and
(\ref{eq:calK}), respectively. 

For a comparison with the SSD expression for 
${\cal M}^{2\nu\varepsilon\beta^{+}}_{J^{\pi}}$ in Eq. (\ref{eq:facMSSDb})
we present its standard form in Table \ref{tab:FacHSD}, which includes 
transitions through the excited states of intermediate nucleus and
is based on the conventional approximation for the  lepton energies
in the energy denominators
\begin{equation}
  \varepsilon_1 \simeq  
  \nu_1 \simeq
  \nu_2 \simeq \frac{w_0^{(2\nu\varepsilon\beta^{+})}}{3},
~~~
    w_0^{(2\nu\varepsilon\beta^{+} )} = \varepsilon_i - \varepsilon_f +
    \varepsilon_{e_1}.
\label{aver2}
\end{equation}

Returning to Eq. (\ref{eq:2vbbwidthb}), we write down the total 
$2\nu\varepsilon\beta^{+}$-decay rate within the SSD hypothesis as
\begin{equation}
\omega^{2\nu\varepsilon\beta^{+}}_{J^\pi} = 
\frac{ln(2)}{
T^{2\nu\varepsilon\beta^{+}}(J^{\pi})}
= c_{2\nu}^{SSD} N^{2\nu\varepsilon\beta^{+}}
  \int\limits_{1}^{w_0^{(\alpha)}} d \varepsilon_1
  \varepsilon_1 \pi_1
  \int\limits_{0}^{w_0^{(\alpha)} - \varepsilon_1} 
  d \nu_1 \nu_1^2 \nu_2^2 a(\varepsilon_1) 
  {\cal K}_{J^{\pi}}(\varepsilon_1, -\varepsilon_{e_1}, \nu_1, \nu_2),
\end{equation}
where  $\alpha = 2\nu\varepsilon\beta^{+}$.

\subsection{The $2\nu\varepsilon\varepsilon$-decay half-life}

The double electron capture differential rate is given by the 
formula~\cite{doi92}
\begin{equation}
  d \omega^{2\nu\varepsilon\varepsilon}_{J^{\pi}} = a_{2\nu} 
N^{2\nu\varepsilon\varepsilon}
    {\cal M}^{2\nu\varepsilon\varepsilon}_{J^{\pi}} 
(-\varepsilon_{e_1}, -\varepsilon_{e_2}, \nu_1, \nu_2)
 d \Omega
  \label{eq:2vbbwidthc}
\end{equation}
where
\begin{equation}
  \label{eq:FacNb}
    N^{2\nu\varepsilon\varepsilon} = 
    \left\{
    \begin{array}{ll}
      (2 \pi^2)^2 \mathfrak{N}_{0,-1}^2& \text{$KK$ capture}\\
      2 (2 \pi^2)^2 \mathfrak{N}_{0,-1} \mathfrak{N}_{1,-1}& \text{$KL_{I}$ 
      capture}
    \end{array}
    \right.,
\end{equation}
\begin{equation}
  d\Omega = 2 \nu_1^2 \nu_2^2 
  \delta(- \varepsilon_{e_1} - \varepsilon_{e_2} + \nu_1 + \nu_2 - 
  \varepsilon_i + \varepsilon_f) 
  d \nu_1 d \nu_2.
\end{equation}
The probability factors $\mathfrak{N}_{0,-1}$, $\mathfrak{N}_{1,-1}$ 
have been defined in the previous subsection.
Their numerical values for some nuclei are shown in Table \ref{tab:eedata}.

Within the SSD hypothesis the matrix element 
$    {\cal M}^{2\nu\varepsilon\varepsilon}_{J^{\pi}} 
(-\varepsilon_{e_1}, -\varepsilon_{e_2}, \nu_1, \nu_2)$ 
in Eq. (\ref{eq:2vbbwidthc}) takes the form
\begin{equation}
  \label{eq:facMSSDc}
    {\cal M}^{2\nu\varepsilon\varepsilon}_{J^{\pi}} 
(-\varepsilon_{e_1}, -\varepsilon_{e_2}, \nu_1, \nu_2)=
    M^{(+)}_{GT, \mathrm{g.s.}}(J^{\pi}) 
    {\cal K}_{J^{\pi}}
    (-\varepsilon_{e_1}, -\varepsilon_{e_1}, \nu_1, \nu_2).
\end{equation}
Here,  $\varepsilon_{e_1}$, $\varepsilon_{e_2}$ are the energies of bound
electrons captured from the $K$ or $L_{I}$ atomic shell. 
The nuclear matrix element and the energy denominator in the r.h.s. of this 
equation are defined in Eqs. (\ref{eq:MEL}) and
(\ref{eq:calK}), respectively. 

In the standard treatment of the $2\nu\varepsilon\varepsilon$-decay
the nuclear matrix element (see in Table \ref{tab:FacHSD})
includes the contribution from transitions through the excited states of
intermediate nucleus, which are constructed within an 
appropriate nuclear model. It is also assumed that
\begin{equation}
  \nu_1 \simeq
  \nu_2 \simeq \frac{w_0^{(2\nu\varepsilon\varepsilon)}}{2},
~~~
    w_0^{(2\nu\varepsilon\varepsilon)} = \varepsilon_i - \varepsilon_f +
    \varepsilon_{e_1} + \varepsilon_{e_2}.
\label{aver3}
\end{equation}
This approximation allows one to separate the phase space integration from 
the evaluation of the $2\nu\varepsilon\varepsilon$-decay 
nuclear matrix element. 

In our derivation of the SSD predictions we always take into account 
the dependence of 
the corresponding nuclear matrix elements on the lepton energies as 
in Eq.~(\ref{eq:facMSSDc}). Thus, the SSD expression for the total 
$2\nu\varepsilon\varepsilon$-decay rate derived from 
Eqs. (\ref{eq:2vbbwidthc}), (\ref{eq:facMSSDc}) reads
\begin{equation}
\omega^{2\nu\varepsilon\varepsilon}_{J^\pi} = 
\frac{ln(2)}{
T^{2\nu\varepsilon\varepsilon}(J^{\pi})}
= c_{2\nu}^{SSD} N^{2\nu\varepsilon\varepsilon}
  \int\limits_{0}^{w_0^{(\alpha)}} 
  d \nu_1 \nu_1^2 \nu_2^2 
  {\cal K}_{J^{\pi}}(-\varepsilon_{e_1}, -\varepsilon_{e_2}, \nu_1, \nu_2)
\end{equation}
with  $\alpha = 2\nu\varepsilon\varepsilon$.

\subsection{Energy distributions of outgoing electrons }

There is a unique possibility to prove or rule out the SSD hypothesis 
for $\beta\beta$-decays by experimental measurement of the 
energy distributions of outgoing electrons \cite{sim01}.
Any deviation from the SSD prediction for the shape of 
the electron energy spectra can be interpreted as 
the influence of the transitions through excited states of intermediate 
nucleus.
The HSD hypothesis, mentioned in Section \ref{sec:theory}, 
reflects an opposite situation to the 
SSD one. It corresponds to the approximation when
one neglects the dependence of the relevant nuclear matrix element
on the lepton energies which is well justified if the
dominant contributions to the $\beta\beta$-decay rate 
come from virtual single-$\beta$-decay transitions via the higher
lying $1^+$ states of intermediate nucleus. 
In this case one substitutes the lepton energies in the energy 
denominators of nuclear matrix elements with their average values 
defined in a special way as in Eqs. (\ref{aver1}), (\ref{aver2}), 
(\ref{aver3}).

For the experimental verification of the SSD and HSD hypotheses the most 
appropriate nuclear systems are those, which exhibit a maximal difference in 
the shapes of the electron/positron energy spectra predicted by the SSD 
and HSD based scenarios, meanwhile having relatively low 
$\beta\beta$-decay half-lives.

Thus, the subject of our interest are the differential rates normalized 
to the total decay rates, namely  
\begin{equation}
  \label{eq:eneprob}
  P^{(\alpha), I}_{J^{\pi}}({\cal E}) = 
  \frac{1}{\omega^{I}_{J^{\pi}}}
  \frac{d \omega^{I}_{J^{\pi}}}{d {\cal E}}
~~~(I=SSD, ~HSD).
\end{equation}
For $2\nu\beta^{\pm}\beta^{\pm}$-decay ($\alpha = 2\nu\beta^{\pm}\beta^{\pm}$)
${\cal E}$ stands for the single electron/positron energy $\varepsilon_1$
or for the sum of the kinetic energies of outgoing 
electrons/positrons  $\tau$. 
For $2\nu\varepsilon\beta^{+}$-decay ($\alpha = 2\nu\varepsilon\beta^{+}$) 
${\cal E}$ denotes the  positron energy  $\varepsilon_1$. 
It is worthwhile to notice that the quantity  
$P^{(\alpha), I}_{J^{\pi}}({\cal E})$ 
does not include any nuclear matrix element because 
both the differential and total rates are scaled with the
same nuclear matrix element which cancels out in their ratio.
The only nuclear structure inputs are the energy release in the 
considered process and the energy difference between  
$1^+$ ground state of intermediate nucleus and parent nucleus. 

We proceed with presenting the expressions for the single electron 
distribution ${d \omega^{I}_{J^{\pi}}}/{d {\varepsilon_1}}$ and  
the summed electron spectrum  ${d \omega^{I}_{J^{\pi}}}/{d \tau }$
for different $\beta\beta$-decay modes within the SSD and HSD hypotheses.  
We consider experimentally most favored ground-state transitions.

{\em The $2\nu\beta^{-}\beta^{-}$-decay to ground state:} \\
i) The single electron differential rate 
\begin{eqnarray}
  \label{2nbb:S}
  \frac{d \omega^{2\nu\beta^{-}\beta^{-}}_{0^+_{g.s.}}}{d \varepsilon_1} &=&
  2 a_{2\nu} |M^{-}_{GT,\mathrm{g.s.}}|^2 \varepsilon_1 \pi_1
  \int\limits_{1}^{w_0^{(\alpha)} - \varepsilon_1} 
  d \varepsilon_2 \varepsilon_2 \pi_2
  \int\limits_{0}^{w_0^{(\alpha)} - \varepsilon_1 -
  \varepsilon_2} 
  d \nu_1 \nu_1^2 \nu_2^2 a(\varepsilon_1,  \varepsilon_2) 
  {\cal K}_{0^+_{g.s.}}(\varepsilon_1, \varepsilon_2, \nu_1, \nu_2)
~~~(SSD), \\
  \label{2nbb:H}
&=&
  2 a_{2\nu}   |{\cal M}_{0^+_{g.s.}}^{\mathrm{HSD}}|^2
  \varepsilon_1 \pi_1
  \int\limits_{1}^{w_0^{(\alpha)} - \varepsilon_1} 
  d \varepsilon_2 \varepsilon_2 \pi_2
  \int\limits_{0}^{w_0^{(\alpha)} - \varepsilon_1 -
  \varepsilon_2} 
  d \nu_1 \nu_1^2 \nu_2^2 a(\varepsilon_1,  \varepsilon_2)  
~~~(HSD).
\end{eqnarray}
Here $\alpha = 2\nu\beta^{-}\beta^{-}$.
We stress again that the corresponding SSD and HSD total decay rates 
are proportional to the squared nuclear matrix elements   
$M^{\pm}_{GT,\mathrm{g.s.}}$ and 
${\cal M}_{0^+_{g.s.}}^{\mathrm{HSD}}$, respectively.
Therefore, they drop out from the normalized differential rates 
$P^{(\alpha), I}_{J^{\pi}}({\varepsilon_1})$ in Eq. (\ref{eq:eneprob}).\\  
ii) The summed electron differential rate
\begin{eqnarray}
  \label{2nbbT:S}
  \frac{d \omega^{2\nu\beta^{-}\beta^{-}}_{0^+_{g.s.}}}{d \tau} &=&
  2 a_{2\nu} |M^{-}_{GT,\mathrm{g.s.}}|^2
  \int\limits_{1}^{\tau + 1} 
  d \varepsilon_1 \varepsilon_1 \pi_1
  \int\limits_{0}^{w_0^{(\alpha)} - \tau - 2}
  d \nu_1 \nu_1^2 \nu_2^2 a(\varepsilon_1,  \varepsilon_2) 
  {\cal K}_{0^+_{g.s.}}(\varepsilon_1, \varepsilon_2, \nu_1, \nu_2),
~~~(SSD),\\
  \label{2nbbT:H}
&=&
  2 a_{2\nu}  |{\cal M}_{0^+_{g.s.}}^{\mathrm{HSD}}|^2
  \int\limits_{1}^{\tau + 1} 
  d \varepsilon_1 \varepsilon_1 \pi_1
  \int\limits_{0}^{w_0^{(\alpha)} - \tau - 2}
  d \nu_1 \nu_1^2 \nu_2^2 a(\varepsilon_1,  \varepsilon_2)
~~~(HSD).
\end{eqnarray}
Here, $\alpha = 2\nu\beta^{-}\beta^{-}$ and 
$\tau = \varepsilon_1+\varepsilon_2-2$ is the sum of the kinetic
energy of emitted electrons in unit of $m_e$.

{\em The $2\nu\varepsilon\beta^{+}$-decay:}\\
The single positron differential rate
\begin{eqnarray}
  \label{2nbbP:S}
  \frac{d \omega^{2\nu\varepsilon\beta^{+}}_{J^{\pi}}}{\varepsilon_1} &=& 
  2 a_{2\nu} |M^{+}_{GT,\mathrm{g.s.}}|^2 N
  \varepsilon_1 \pi_1
  \int\limits_{0}^{w_0^{(\alpha)} - \varepsilon_1} 
  d \nu_1 \nu_1^2 \nu_2^2 a(\varepsilon_1) 
  {\cal K}_{J^{\pi}}(\varepsilon_1, -\varepsilon_{e_1}, \nu_1, \nu_2)
~~~(SSD),\\
  \label{2nbbP:H}
&=&
  2 a_{2\nu} N |{\cal M}_{J^{\pi}}^{\mathrm{HSD}}|^2
  \varepsilon_1 \pi_1
  \int\limits_{0}^{w_0^{(\alpha)} - \varepsilon_1} 
  d \nu_1 \nu_1^2 \nu_2^2 a(\varepsilon_1)
~~~(HSD).
\end{eqnarray}
Here $\alpha = 2\nu\varepsilon\beta^{+}$.

In this paper we do not discuss the single positron energy distribution
and summed positron energy spectrum of the $2\nu\beta^{+}\beta^{+}$-decay. 
As we commented in Section \ref{sec:theory},  this process is less favorable 
for experimental study than the other $\beta\beta$-decay modes 
because of its small available energy and the Coulomb suppression.
Also we do not discuss the angular distributions in 
$2\nu\beta^{\pm}\beta^{\pm}$-decay \cite{sim01},
which may also be helpful for the experimental verification of the 
SSD and HSD hypotheses.

There may happen the situation when 
the measured energy distributions do not correspond neither to the SSD nor HSD 
predictions, so that the SSD is only partially realized and there are 
interfering contributions from the higher-lying $1^+$ 
states of intermediate nucleus. 
In this case one may apply to the data analysis the approximate 
expression for the mixed SSD-HSD distribution:
\begin{equation}
  P^{(\alpha), exp.}_{J^{\pi}}({\cal E}) \approx
 x~ P^{(\alpha), SSD}_{J^{\pi}}({\cal E}) +
(1-x)~  P^{(\alpha), HSD}_{J^{\pi}}({\cal E}) 
~~~(0\le x \le 1),
\label{mix}
\end{equation}
where the parameter $x$ characterizes the relative contribution 
from the first $1^+$ state of the double-odd nucleus to the corresponding 
mode of $\beta\beta$-decay. Fitting the above formula to the experimental 
distributions one can extract the value of $x$ and get an idea on the role 
of this contributions in the studied process. 
However, one should take into account that the accurate definition 
of the mixed distributions in Eq. (\ref{mix}) 
requires the knowledge of nuclear matrix elements involved in the 
differential rates.

\section{\label{sec:res}Results and discussion}

In the present work we have analyzed the neutrino accompanied 
double beta decay modes for emitters, which fulfil the SSD hypothesis 
condition: The ground state of odd-odd nucleus must be $1^+$ state.
The subject of our interest are eight even-even nuclei, which satisfy this
condition  for the $2\nu\beta^{-}\beta^{-}$-decay,
and ten double electron capture unstable nuclei.
Their basic parameters are listed in Tables \ref{tab:bbdata} 
and \ref{tab:eedata}.
>From Table \ref{tab:bbdata} one may notice that in the case 
of $^{100}Mo$ and $^{116}Cd$ the energy difference $\Delta$ between 
the ground states of initial and intermediate nuclei is negative. 
Therefore for these nuclear systems  the difference between the maximal
and minimal values of the energy denominators of $\beta\beta$-nuclear
matrix elements is significant. 
In this case it is especially important to properly take into account the
dependence of nuclear matrix elements on lepton energies in the
calculation of decay rate. 

In Table \ref{tab:eedata} we listed the dimensionless 
normalization constants $\mathfrak{N}_{0,-1}$ and  
$\mathfrak{N}_{1,-1}$~\cite{doi92} of relativistic wave function 
of bound electron  in the $K$ ($1s_{1/2}$) and $L_I$
($2s_{1/2}$) states, respectively. Comparing the values of these 
constants one can see that the electron capture from $L_I$ shell 
is suppressed by about a
factor of 10 in comparison with the electron capture from $K$ shell. 
The energies of bound electron and normalization quantities 
$\mathfrak{N}_{0,-1}$ and  $\mathfrak{N}_{1,-1}$ were
calculated assuming the Coulomb potential for finite-size spherical nucleus 
($R = r_0 A^{1/3}$, $r_0 = 1.2~\mathrm{fm}$) with uniform charge distribution. 

We calculated the half-lives for the two-neutrino double beta decays 
of the nuclei listed in Tables \ref{tab:bbdata} and \ref{tab:eedata}
within the SSD hypothesis. 
For this purpose we used the $ft$-values corresponding to the single 
$\beta^-$ and electron-capture
transitions feeding the low-lying states of initial and final nuclei.
They were taken from NNDC On Line Data Service of ENSDF database 
~\cite{bhat92}. For those nuclear systems, where only the lower limits
on the $log ft$-value is known, we derived the lower limits on the 
$\beta\beta$-decay half-lives.  
The calculation of $\beta\beta$-decay rates has been
performed with the exact treatment of the energy denominators 
of perturbation theory. Transitions to the final $0^+_{g.s.}$ ground and 
$0^+$ and $2^+$ excited states have been considered.  
In the case of  $2\nu\varepsilon\beta^{+}$-decay and 
$2\nu\varepsilon\varepsilon$-decay modes 
the captures from $K$ and $L_I$ shells have been taken into account.
We neglected the captures from $L_{II}$ ($2p_{1/2}$)  and $L_{III}$ 
($2p_{3/2}$) shells suppressed by the factor $(r/R)$ in comparison 
with s-states.

We improved our previous results ~\cite{sem00,sim01} by more
accurate treatment of the Coulomb correction factor 
$R_{11}^{\pm}(\varepsilon)$ [see Eqs. (\ref{ecoulf}) and (\ref{eq:R11})], 
which reflects the charge distribution 
of the final nucleus. 
We have found that the presence of this factor affects
the $2\nu\beta^{-}\beta^{-}$-decay half-life only slightly by
reducing its value up to $5 - 10\%$.
On the other hand, it plays an important role in the evaluation
of the $2\nu\beta^{+}\beta^{+}$-decay half-lives enhancing their values 
by $20 - 50\%$ (Tab.~\ref{tab:bbthalfr11}).

The results for the $2\nu\beta^-\beta^-$-transitions to the final $0^+_{g.s.}$ 
ground and $0^+$, $2^+$  excited states are shown in Table \ref{tab:bbhalf}.
It is seen that in the case of $^{100}Mo$ the SSD hypothesis predictions are 
consistent with the experimental half-lives both for the transitions
to ground and $0^+_{1}$ excited states. On the other hand for $^{116}Cd$ 
the calculated half-life to the final ground state is by about a 
factor 2 shorter than the experimental value. A similar situation
occurs also for the $2\nu\beta^-\beta^-$-decay of $^{128}Te$.
One possible explanation of this discrepancy may consist in 
the cancellation between the transitions
passing through the $1^+$ ground state  of double odd nuclei and the
rest of the transitions going through the excited $1^+$ states. However, 
one should also keep in mind 
the uncertainties both in the measured $2\nu\beta^-\beta^-$-decay half-life 
and in the experimental determination of the $log ft$ value 
of electron capture.
In view of these uncertainties it is not possible to make definite 
conclusion on the validity of the SSD hypothesis for the case of this nucleus. 
For other $2\nu\beta^-\beta^-$-transitions 
in  Table \ref{tab:bbhalf} there are 
only lower limits on the half-life available.

In these cases the SSD hypothesis helps us to give an order 
of magnitude estimates for the the $2\nu\beta^-\beta^-$-decay half-life. 
One may note that the $2\nu\beta^-\beta^-$-decays of $^{100}Mo$ and $^{116}Cd$ 
to excited $0^+_1$ state are significantly faster 
than those to the $2^+_1$ state in spite of the smaller energy
release. The suppression of the $2^+$ decay channel is due to
the mutual cancellation of $K$ and $L$  terms in the
$2\nu\beta^-\beta^-$-decay amplitude [see Eq. (\ref{eq:calK})].

The results for the $2\nu\beta^+\beta^+$-, $2\nu\varepsilon\beta^{+}$- and
$2\nu\varepsilon\varepsilon$-decays are listed in Tables 
~\ref{tab:eehalf}, ~\ref{tab:eechalf} and ~\ref{tab:ecechalf}.
So far none of these $\beta\beta$-decay modes have been seen experimentally.
The SSD hypothesis calculations of decay rates require the knowledge
of the measured $log ft$-values for single $\beta^-$ and electron-capture
decays. However, the $log ft$-values for the $\beta^-$ decay of 
the ground state of odd-odd nucleus are not known for 
$A = 120, 136, 162$ systems and, therefore, we do not present 
the SSD predictions for the half-lives of the corresponding processes.
In the case of ${^{164}\mathrm{Er}}$ its small available energy
allows only $2\nu\varepsilon\varepsilon$ mode, which is
strongly suppressed. Therefore, the SSD predictions for this transition are
not presented. In particular cases, only lower limits on the 
$log ft$ value for the single $\beta^-$-decay of the ground
state of intermediate nucleus are measured. This allows us to
determine only the lower limits on the corresponding half-lives
within the SSD based approach. At present the
attention of experimentalists is concentrated on the
$2\nu\varepsilon\varepsilon$-decay of $^{106}Cd$ \cite{TGV}. 
For the ground state to ground state transition of
this  process the SSD hypothesis predicts the half-life larger than
$4.4\times 10^{21}$ years.

\begin{table}
  \caption{%
The effect of the correction factor $R_{11}^{-}(\varepsilon)$ 
(see Eq.~\ref{eq:R11})
on the calculation of $2\nu\beta^{\pm}\beta^{\pm}$-decay rate. The difference
between the results with and without the correction factor is given in
percents. Only transitions to the ground state of final nucleus are
considered.
 \label{tab:bbthalfr11}}
  \begin{center}
\begin{tabular}{cccccccc|cccc}
  \hline
  \hline
${^{70}\mathrm{Zn}}$
&${^{80}\mathrm{Se}}$
&${^{100}\mathrm{Mo}}$
&${^{104}\mathrm{Ru}}$
&${^{110}\mathrm{Pd}}$
&${^{114}\mathrm{Cd}}$
&${^{116}\mathrm{Cd}}$
&${^{128}\mathrm{Te}}$ 
&${^{78}\mathrm{Kr}}$
&${^{106}\mathrm{Cd}}$
&${^{130}\mathrm{Ba}}$
&${^{136}\mathrm{Ce}}$\\
 \hline
$2.6$&
$3.5$&
$4.6$&
$5.7$&
$6.0$&
$7.0$&
$6.2$&
$8.1$& 
$19$&
$33$&
$42$&
$42$\\ 
\hline
\hline
\end{tabular}
\end{center}
\end{table}

\begin{table}[!ht]
  \caption{%
The half-lives   ($T^{(2\nu\beta^-\beta^-)}_{\mathrm{SSD}}$) 
for the $2\nu\beta^-\beta^-$-decay transitions 
to the ground state ($J^\pi=0^+_{g.s.}$) and excited states ($J^\pi=0^+$ and $2^+$) calculated
within the SSD hypothesis. The excitation energies
of the states, $E^{exc}$, and their 
$\log {ft_i}$ (electron capture) and $\log {ft_f}$ (single $\beta^-$-decay) 
are shown.
The corresponding values were taken from NNDC On Line Data Service 
from ENSDF database updated on
 08/13/2003~\cite{bhat92}. Here $W_0 = m_e w_0^{(2\nu\beta^-\beta^-)}$ 
(see Eq.~(\ref{aver1})).  The experimental $2\nu\beta\beta$-decay 
half-lives $T^{(2\nu\beta^-\beta^-)}_{\mathrm{exp}}$ for
$^{100}Mo$ to ground and excited $0^+_1$ states are from
Refs. \protect\cite{mo100g} and \protect\cite{aver}, respectively.
The $2\nu\beta\beta$-decay half-life of $^{116}Cd$ is from Ref.
\cite{cd116}. For the lower limit on
$T^{(2\nu\beta^-\beta^-)}_{\mathrm{exp}}$
we used the best limit presented in Tables of Ref. \cite{data},
which is a compilation of experimental data. 
  \label{tab:bbhalf}}
  \begin{center}
    \begin{tabular}{cc}
\begin{tabular}{lllllll}
\hline
\hline
    Nucl.&$J^{\pi}_{f}$&$E^{\mathrm{exc}}$&
    $\log ft_{f}$&$W_0$&$T_{\mathrm{SSD}}^{(2\nu\beta^-\beta^-)}$
    &$T_{\mathrm{exp}}^{(2\nu\beta^-\beta^-)}$\\ 
$\log ft_{i}$&&(keV)&&(MeV)&(years)&(years)\\ 
\hline 
${^{70}\mathrm{Zn}}$&$0^+_{\mathrm{g.s.}}$&0.0&$5.1$&$2.0229$&$7.0~10^{23}$&--\\ 
$4.725$&&&&&&\\ 
${^{80}\mathrm{Se}}$&$0^+_{\mathrm{g.s.}}$&0.0&$5.484$&$1.1559$&$9.1~10^{30}$&
--\\ 
$4.67$&&&&&&\\ 
${^{100}\mathrm{Mo}}$&$0^+_{\mathrm{g.s.}}$&0.0&$4.6$&$4.0563$&$7.3~10^{18}$&$6.8~10^{18}$\\ 
$4.45$&$2^+_{1}$&539.59&$6.5$&$3.5167$&$1.7~10^{23}$&$>1.6~10^{21}$\\ 
&$0^+_{1}$&1130.42&$5.0$&$2.9259$&$4.2~10^{20}$&$6.1~10^{20}$\\ 
&$2^+_{2}$&1362.25&$7.1$&$2.6941$&$1.4~10^{25}$&$>1.3~10^{21}$\\ 
&$0^+_{2}$&1740.8&$6.3$&$2.3155$&$1.1~10^{23}$&$>1.3~10^{21}$\\ 
&$2^+_{3}$&1865.2&$6.5$&$2.1911$&$5.5~10^{25}$&--\\ 
${^{104}\mathrm{Ru}}$&$0^+_{\mathrm{g.s.}}$&0.0&$4.55$&$2.3216$&$6.4~10^{21}$&--\\ 
$4.32$&$2^+_{1}$&555.81&$5.8$&$1.7658$&$1.8~10^{29}$&--\\ 
${^{110}\mathrm{Pd}}$&$0^+_{\mathrm{g.s.}}$&0.0&$4.66$&$3.0217$&$1.2~10^{20}$&$>6.0~10^{16}$\\ 
$4.08$&$2^+_{1}$&657.52&$5.528$&$2.3642$&$4.4~10^{25}$&--\\ 
&&&&&&\\ 
\hline
\hline
\end{tabular}&
    \begin{tabular}{lllllll}
\hline
\hline
    Nucl.&$J^{\pi}_{f}$&$E^{\mathrm{exc}}$&
    $\log ft_{f}$&$W_0$&$T_{\mathrm{SSD}}^{(2\nu\beta^-\beta^-)}$&
    $T_{\mathrm{exp}}^{(2\nu\beta^-\beta^-)}$\\ 
$\log ft_{i}$&&(keV)&&(MeV)&(years)&(years)\\ 
\hline 
${^{110}\mathrm{Pd}}$&$0^+_{1}$&1473.03&$6.8$&$1.5487$&$2.4~10^{26}$&--\\ 
$4.08$&$2^+_{2}$&1475.74&$7.39$&$1.5460$&$3.8~10^{31}$&--\\ 
&$0^+_{2}$&1731.53&$8.1$&$1.2902$&$5.3~10^{29}$&--\\ 
&$2^+_{3}$&1783.35&$6.9$&$1.2384$&$1.3~10^{35}$&--\\ 
${^{114}\mathrm{Cd}}$&$0^+_{\mathrm{g.s.}}$&0.0&$4.473$&$1.5588$&$1.1~10^{25}$&$>9.2~10^{16}$\\ 
$4.89$&&&&&&\\ 
${^{116}\mathrm{Cd}}$&$0^+_{\mathrm{g.s.}}$&0.0&$4.662$&$3.8270$&$1.1~10^{19}$&$2.6~10^{19}$\\ 
$4.39$&$2^+_{1}$&1293.4&$5.85$&$2.5336$&$6.8~10^{24}$&$>2.4~10^{21}$\\ 
&$0^+_{1}$&1756.8&$5.88$&$2.0702$&$1.8~10^{23}$&$>2.1~10^{21}$\\ 
&$2^+_{2}$&2112.1&$6.31$&$1.7149$&$2.3~10^{28}$&$>1.7~10^{20}$\\ 
&$2^+_{3}$&2225.5&$6.4$&$1.6015$&$1.5~10^{29}$&$>1.0~10^{20}$\\ 
&$0^+_{2}$&2546.0&$5.99$&$1.2810$&$3.0~10^{27}$&--\\ 
&$2^+_{4}$&2649.8&$5.79$&$1.1772$&$1.8~10^{34}$&--\\ 
${^{128}\mathrm{Te}}$&$0^+_{\mathrm{g.s.}}$&0.0&$6.061$&$1.8892$&$1.1~10^{25}$&$2.2~10^{24}$\\ 
$5.049$&$2^+_{1}$&442.901&$6.495$&$1.4463$&$1.2~10^{33}$&$>4.7~10^{21}$\\ 
\hline
\hline
\end{tabular}
\end{tabular}
\end{center}
\end{table}

\begin{table}[!ht]
  \caption{%
The half-lives  ($T^{(2\nu\beta^+\beta^+)}_{\mathrm{SSD}}$) 
for the $2\nu\beta^+\beta^+$-decay transitions 
to the ground state ($J^\pi=0^+_{g.s.}$) and excited states 
($J^\pi=0^+$ and $2^+$) calculated within the SSD hypothesis. 
$\log_{ft_i}$ and $\log_{ft_f}$  denote $log ft$-values for 
single $\beta^-$- and electron capture-decays, respectively. 
Here $W_0 = m_e w_0^{(2\nu\beta^+\beta^+)}$ (see Eq.~(\ref{aver1})).  
For other notations see Tab~\ref{tab:bbhalf}.
  \label{tab:eehalf}}
  \begin{center}
    \begin{tabular}{cc}
\begin{tabular}{lllllll}
\hline
\hline
    Nucl.&$J^{\pi}_{f}$&$E^{\mathrm{exc}}$&$\log
    ft_{f}$&$W_0$&$T_{\mathrm{SSD}}^{(2\nu\beta^+\beta^+)}$&
    $T_{\mathrm{exp}}^{(2\nu\beta^+\beta^+)}$\\ 
$\log ft_{i}$&&(keV)&&(MeV)&(years)&(years)\\ 
\hline 
${^{78}\mathrm{Kr}}$\footnotemark[1]&$0^+_{\mathrm{g.s.}}$&0.0&$4.752$&$1.8440$&$>1.2~10^{28}$&$>2.0~10^{21}$\\ 
$>5.6$&$2^+_{1}$&613.71&$5.069$&$1.2303$&$>4.6~10^{41}$&--\\
&&&&&\\
&&&&&\\
\hline
\hline
\end{tabular}&
\begin{tabular}{lllllll}
\hline
\hline
    Nucl.&$J^{\pi}_{f}$&$E^{\mathrm{exc}}$&$\log
    ft_{f}$&$W_0$&$T_{\mathrm{SSD}}^{(2\nu\beta^+\beta^+)}$&
    $T_{\mathrm{exp}}^{(2\nu\beta^+\beta^+)}$\\ 
$\log ft_{i}$&&(keV)&&(MeV)&(years)&(years)\\ 
\hline 
${^{106}\mathrm{Cd}}$\footnotemark[1]&$0^+_{\mathrm{g.s.}}$&0.0&$4.92$&$1.7491$&$>2.4~10^{27}$&
$>2.4~10^{20}$\\ 
$>4.1$&$2^+_{1}$&511.85&$5.24$&$1.2373$&$>1.1~10^{40}$&$>1.6~10^{20}$\\ 
${^{130}\mathrm{Ba}}$&$0^+_{\mathrm{g.s.}}$&0.0&$5.073$&$1.5886$&$1.7~10^{30}$&$>4.0~10^{21}$\\ 
$5.36$&$2^+_{1}$&536.95&$6.3$&$1.0516$&$1.2~10^{60}$&$>4.0~10^{21}$\\ 
\hline
\hline
\end{tabular}
\end{tabular}
\footnotetext[1]{Only the lower limits of half-lives are calculated}
\end{center}
\end{table}

\begin{table}[!ht]
  \caption{%
The half-lives    ($T^{(2\nu\varepsilon\beta^{+})}_{\mathrm{SSD}}$) 
for the $2\nu\varepsilon\beta^{+}$-decay transitions 
to the ground state ($J^\pi=0^+_{g.s.}$) and excited states ($J^\pi=0^+$ and $2^+$) calculated 
within the SSD hypothesis. 
Here $W_0 = m_e w_0^{(2\nu\varepsilon\beta^{+})}$ (see Eq.~(\ref{aver2})).  
The values in the first and the second rows for each transition
correspond to the case of electron capture from the $K$
and $L_{I}$ atomic shells, respectively.
Other notations are the same as in Tab~\ref{tab:bbhalf}. 
  \label{tab:eechalf}}
  \begin{center}
    \begin{tabular}{cc}
\begin{tabular}{lllllll}
\hline
\hline
    Nucl.&$J^{\pi}_{f}$&$E^{\mathrm{exc}}$&$\log
    ft_{f}$&$W_0$&$T_{\mathrm{SSD}}^{(2\nu\varepsilon\beta^{+})}$&
    $T_{\mathrm{exp}}^{(2\nu\varepsilon\beta^{+})}$\\ 
$\log ft_{i}$&&(keV)&&(MeV)&(years)&(years)\\ 
\hline 
${^{64}\mathrm{Zn}}$&$0^+_{\mathrm{g.s.}}$&0.0&$4.973$&$0.5730$&$9.2~10^{34}$&$>2.3~10^{18}$\\ 
$5.293$&&&&$0.5823$&$2.2~10^{35}$&$>2.3~10^{18}$\\ 
${^{78}\mathrm{Kr}}$\footnotemark[1]&$0^+_{\mathrm{g.s.}}$&0.0&$4.752$&$2.3370$&$>2.3~10^{24}$&$>1.1~10^{20}$\\ 
$>5.6$&&&&$2.3505$&$>1.6~10^{25}$&--\\ 
&$2^+_{1}$&613.71&$5.069$&$1.7233$&$>9.0~10^{27}$&--\\ 
&&&&$1.7368$&$>5.8~10^{28}$&--\\ 
&$2^+_{2}$&1308.48&$6.62$&$1.0285$&$>3.6~10^{32}$&--\\ 
&&&&$1.0420$&$>2.0~10^{33}$&--\\ 
&$0^+_{1}$&1498.41&$6.47$&$0.8386$&$>1.6~10^{31}$&--\\ 
&&&&$0.8521$&$>9.0~10^{31}$&--\\ 
&$0^+_{2}$&1758.31&$6.78$&$0.5787$&$>8.1~10^{36}$&--\\ 
&&&&$0.5922$&$>1.3~10^{37}$&--\\ 
${^{106}\mathrm{Cd}}$\footnotemark[1]&$0^+_{\mathrm{g.s.}}$&0.0&$4.92$&$2.2278$&$>2.7~10^{22}$&$>4.1~10^{20}$\\ 
$>4.1$&&&&$2.2520$&$>1.7~10^{23}$&$>4.1~10^{20}$\\ 
&$2^+_{1}$&511.85&$5.24$&$1.7159$&$>1.1~10^{25}$&$>2.6~10^{20}$\\ 
&&&&$1.7401$&$>5.9~10^{25}$&$>2.6~10^{20}$\\ 
\hline
\hline
\end{tabular}&
\begin{tabular}{lllllll}
\hline
\hline
    Nucl.&$J^{\pi}_{f}$&$E^{\mathrm{exc}}$&
    $\log ft_{f}$&$W_0$&$T_{\mathrm{SSD}}^{(2\nu\varepsilon\beta^{+})}$&
    $T_{\mathrm{exp}}^{(2\nu\varepsilon\beta^{+})}$\\ 
$\log ft_{i}$&&(keV)&&(MeV)&(years)&(years)\\ 
\hline 
${^{106}\mathrm{Cd}}$\footnotemark[1]&$0^+_{1}$&1133.77&$6.5$&$1.0940$&$>1.1~10^{27}$&$>1.1~10^{20}$\\ 
$>4.1$&&&&$1.1182$&$>5.5~10^{27}$&$>1.1~10^{20}$\\ 
&$2^+_{3}$&1562.26&$6.5$&$0.6655$&$>5.4~10^{33}$&--\\ 
&&&&$0.6897$&$>6.5~10^{33}$&--\\ 
&$0^+_{2}$&1706.39&$7.0$&$0.5214$&$>4.5~10^{43}$&--\\ 
&&&&$0.5456$&$>2.8~10^{38}$&--\\ 
${^{112}\mathrm{Sn}}$&$0^+_{\mathrm{g.s.}}$&0.0&$4.7$&$1.3760$&$3.8~10^{24}$&--\\ 
$4.12$&&&&$1.4023$&$2.1~10^{25}$&--\\ 
&$2^+_{1}$&617.11&$5.309$&$0.7589$&$2.3~10^{32}$&--\\ 
&&&&$0.7852$&$5.1~10^{32}$&--\\ 
${^{130}\mathrm{Ba}}$&$0^+_{\mathrm{g.s.}}$&0.0&$5.073$&$2.0550$&$1.3~10^{24}$&$>4.0~10^{21}$\\ 
$5.36$&&&&$2.0883$&$7.1~10^{24}$&$>4.0~10^{21}$\\ 
&$2^+_{1}$&536.95&$6.3$&$1.5180$&$2.1~10^{28}$&$>4.0~10^{21}$\\ 
&&&&$1.5514$&$9.2~10^{28}$&$>4.0~10^{21}$\\ 
&$2^+_{2}$&1122.36&$7.5$&$0.9326$&$4.9~10^{32}$&$>4.0~10^{21}$\\ 
&&&&$0.9659$&$1.3~10^{33}$&$>4.0~10^{21}$\\ 
\hline
\hline
\end{tabular}
\end{tabular}
\footnotetext[1]{Only the lower limits on half-lives are calculated}
\end{center}
\end{table}

\begin{table}[!ht]
  \caption{%
The half-lives    ($T^{(2\nu \varepsilon\varepsilon)}_{\mathrm{SSD}}$) 
for the $2\nu \varepsilon\varepsilon$-decay  transitions 
to the ground state ($J^\pi=0^+_{g.s.}$) and excited states ($J^\pi=0^+$ and $2^+$) calculated
within the SSD hypothesis. 
Here $W_0 = m_e w_0^{(2\nu \varepsilon\varepsilon)}$ (see Eq.~(\ref{aver3})).
For transitions to $0^+$ state 
the double capture of atomic electrons 
from K-shell 
(upper row) and from $KL_I$ (lower row) shells  is considered. In the case
of transitions to $2^+$ state only 
the double capture of atomic electrons
from K-shell is taken into account.
Other notations are the same as in Tab~\ref{tab:bbhalf}.
  \label{tab:ecechalf}}
  \begin{center}
    \begin{tabular}{cc}
\begin{tabular}{llllllll}
\hline
\hline
    Nucl.&$J^{\pi}_{f}$&$E^{\mathrm{exc}}$&$\log ft_{f}$&$W_0$&
    $T_{\mathrm{SSD}}^{(2\nu \varepsilon\varepsilon)}$&$T_{\mathrm{exp}}^{(2\nu \varepsilon\varepsilon)}$\\ 
$\log ft_{i}$&&keV&&(MeV)&(years)&(years)\\ 
\hline 
${^{64}\mathrm{Zn}}$&$0^+_{\mathrm{g.s.}}$&0.0&$4.973$&$1.0716$&$1.9~10^{26}$&$>8.0~10^{15}$\\
$5.293$&&&&$1.0809$&$7.1~10^{26}$&$>8.0~10^{15}$\\ 
${^{78}\mathrm{Kr}}$\footnotemark[1]&$0^+_{\mathrm{g.s.}}$&0.0&$4.752$&$2.8301$&$>1.6~10^{24}$&$>2.3~10^{20}$\\ 
$>5.6$&&&&$2.8435$&$>5.9~10^{24}$&--\\ 
&$2^+_{1}$&613.71&$5.069$&$2.2298$&$>5.1~10^{30}$&--\\ 
&$2^+_{2}$&1308.48&$6.62$&$1.5350$&$>8.5~10^{32}$&--\\ 
&$0^+_{1}$&1498.41&$6.47$&$1.3317$&$>1.7~10^{27}$&--\\ 
&&&&$1.3451$&$>6.0~10^{27}$&--\\ 
&$0^+_{2}$&1758.31&$6.78$&$1.0718$&$>8.4~10^{27}$&--\\ 
&&&&$1.0852$&$>3.0~10^{28}$&--\\ 
&$2^+_{3}$&1996.08&$6.93$&$0.8474$&$>2.8~10^{34}$&--\\ 
&$2^+_{4}$&2329.4&$7.37$&$0.5141$&$>1.1~10^{36}$&--\\ 
&$0^+_{3}$&2334.43&$5.9$&$0.4956$&$>3.2~10^{28}$&--\\ 
&&&&$0.5091$&$>1.1~10^{29}$&--\\ 
&$2^+_{5}$&2537.18&$5.61$&$0.3063$&$>3.9~10^{35}$&--\\ 
${^{106}\mathrm{Cd}}$\footnotemark[1]&$0^+_{\mathrm{g.s.}}$&0.0&$4.92$&$2.7064$&$>4.4~10^{21}$&$>5.8~10^{17}$\\ 
$>4.1$&&&&$2.7306$&$>1.5~10^{22}$&$>1.5~10^{17}$\\ 
&$2^+_{1}$&511.85&$5.24$&$2.2187$&$>4.0~10^{26}$&$>3.5~10^{17}$\\ 
&$0^+_{1}$&1133.77&$6.5$&$1.5726$&$>1.1~10^{24}$&$>7.3~10^{19}$\\ 
&&&&$1.5968$&$>3.7~10^{24}$&$>7.3~10^{19}$\\ 
&$2^+_{3}$&1562.26&$6.5$&$1.1683$&$>5.4~10^{28}$&--\\ 
&$0^+_{2}$&1706.39&$7.0$&$1.0000$&$>1.9~10^{25}$&--\\ 
&&&&$1.0242$&$>5.8~10^{25}$&--\\ 
&$2^+_{4}$&1909.47&$7.9$&$0.8211$&$>4.8~10^{30}$&--\\ 
&&&&&&\\
&&&&&&\\
\hline
\hline
\end{tabular}&
\begin{tabular}{llllllll}
\hline
\hline
    Nucl.&$J^{\pi}_{f}$&$E^{\mathrm{exc}}$&$\log ft_{f}$&$W_0$&
    $T_{\mathrm{SSD}}^{(2\nu \varepsilon\varepsilon)}$&$T_{\mathrm{exp}}^{(2\nu \varepsilon\varepsilon)}$\\ 
$\log ft_{i}$&&keV&&(MeV)&(years)&(years)\\ 
\hline 
${^{106}\mathrm{Cd}}$\footnotemark[1]&$0^+_{3}$&2001.49&$6.1$&$0.7049$&$>8.9~10^{24}$&--\\ 
$>4.1$&&&&$0.7291$&$>2.7~10^{25}$&--\\ 
&$2^+_{5}$&2242.53&$6.9$&$0.4881$&$>3.8~10^{30}$&--\\ 
&$0^+_{4}$&2277.86&$7.6$&$0.4285$&$>2.1~10^{27}$&--\\ 
&&&&$0.4527$&$>5.6~10^{27}$&--\\ 
&$2^+_{6}$&2308.86&$6.8$&$0.4217$&$>5.8~10^{30}$&--\\ 
&$2^+_{7}$&2439.22&$7.8$&$0.2914$&$>3.2~10^{32}$&--\\ 
${^{108}\mathrm{Cd}}$&$0^+_{\mathrm{g.s.}}$&0.0&$4.7$&$0.2044$&$3.9~10^{27}$&$>4.1~10^{17}$\\ 
$4.425$&&&&$0.2286$&$8.0~10^{27}$&--\\ 
${^{112}\mathrm{Sn}}$&$0^+_{\mathrm{g.s.}}$&0.0&$4.7$&$1.8518$&$1.7~10^{22}$&--\\ 
$4.12$&&&&$1.8781$&$5.5~10^{22}$&--\\ 
&$2^+_{1}$&617.11&$5.309$&$1.2610$&$4.9~10^{28}$&--\\ 
&$0^+_{1}$&1223.52&$5.376$&$0.6282$&$7.4~10^{24}$&--\\ 
&&&&$0.6546$&$2.1~10^{25}$&--\\ 
&$2^+_{2}$&1312.1&$7.17$&$0.5660$&$1.9~10^{32}$&--\\ 
&$2^+_{3}$&1468.04&$5.9$&$0.4101$&$6.2~10^{31}$&--\\ 
&$0^+_{2}$&1870.21&$5.51$&$0.0079$&$5.4~10^{34}$&--\\ 
${^{130}\mathrm{Ba}}$&$0^+_{\mathrm{g.s.}}$&0.0&$5.073$&$2.5214$&$5.0~10^{22}$&$>4.0~10^{21}$\\ 
$5.36$&&&&$2.5547$&$1.6~10^{23}$&$>4.0~10^{21}$\\ 
&$2^+_{1}$&536.95&$6.3$&$2.0177$&$5.5~10^{28}$&$>4.0~10^{21}$\\ 
&$2^+_{2}$&1122.36&$7.5$&$1.4323$&$3.0~10^{30}$&$>4.0~10^{21}$\\ 
&$0^+_{1}$&1793.7&$7.0$&$0.7277$&$5.3~10^{26}$&$>4.0~10^{21}$\\ 
&&&&$0.7610$&$1.4~10^{27}$&$>4.0~10^{21}$\\ 
&$0^+_{2}$&2017.11&$6.2$&$0.5043$&$3.9~10^{26}$&$>4.0~10^{21}$\\ 
&&&&$0.5376$&$9.8~10^{26}$&$>4.0~10^{21}$\\ 
&$2^+_{3}$&2151.0&$6.2$&$0.4037$&$4.4~10^{31}$&$>4.0~10^{21}$\\ 
\hline
\hline
\end{tabular}
\end{tabular}
\footnotetext[1]{Only the lower limits on half-lives are calculated}
\end{center}
\end{table}

\begin{figure}
\begin{center}
\includegraphics[width=0.6\textwidth,height=0.7\textheight]{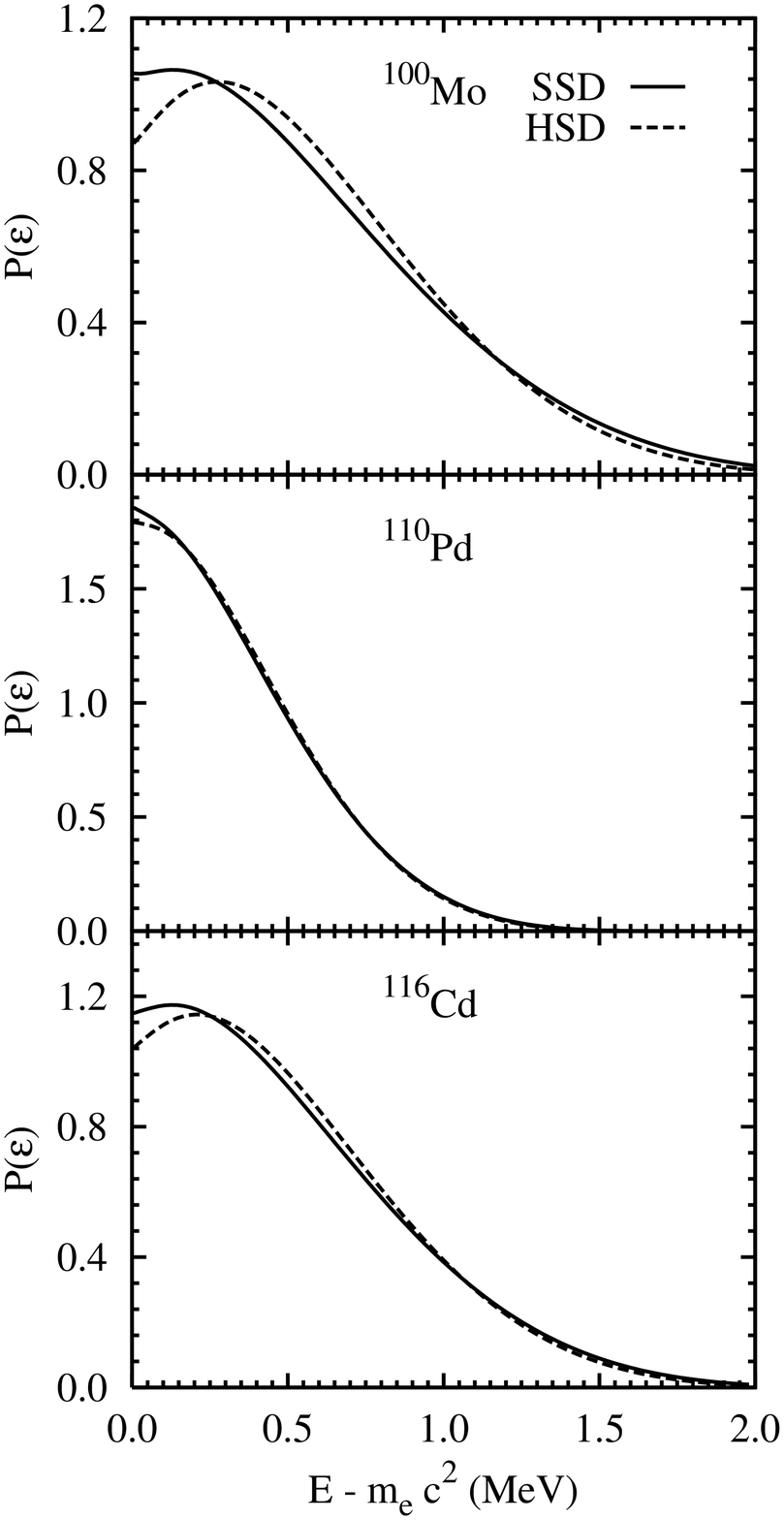}
\end{center}
\caption{
The single electron differential decay rate 
normalized to the total decay rate 
$P(\varepsilon) = (1/\omega) d \omega/d \varepsilon $ vs.
the electron energy $\varepsilon$ 
for $2\nu\beta^-\beta^-$-decay to the $0^+$ ground state.
The results are presented for the cases of 
${^{100}\mathrm{Mo}}\rightarrow {^{100}\mathrm{Ru}}$, 
${^{110}\mathrm{Pd}}\rightarrow {^{110}\mathrm{Cd}}$, 
and 
${^{116}\mathrm{Cd}}\rightarrow {^{116}\mathrm{Sn}}$.
The calculations have been performed within the single-state 
dominance hypothesis
(SSD) and with the assumption of dominance of higher lying states (HSD). 
\label{fig:gs2gs}}
\end{figure}

In Fig.~\ref{fig:gs2gs} we present the single electron differential decay
rate   $P^{(\alpha), I}_{J^{\pi}}({\cal E})$ [see
 Eqs. (\ref{eq:eneprob}), (\ref{2nbb:S}) and (\ref{2nbb:S})] for
the $2\nu\beta^{-}\beta^{-}$-decay of $^{100}Mo$, $^{110}Pd$ and
$^{116}Mo$ to the $0^+$ ground state. We see that there is a notable 
difference in the
behavior of the single electron energy distribution calculated 
within the SSD and HSD approaches especially for the small electron
energies. The effect is significant in the case of $^{100}Mo$ and
$^{116}Cd$. This is due to the smallness of the energy difference between the
ground states of the intermediate and initial nuclei in comparison with 
the energy release in the $2\nu\beta^{-}\beta^{-}$-decay 
process (see Table \ref{tab:bbdata}). We also note that 
the $2\nu\beta^{-}\beta^{-}$-decay of $^{100}Mo$ is slightly more
favored for the experimental verification of the SSD hypothesis 
than the  $2\nu\beta^{-}\beta^{-}$-decay of $^{116}Cd$
due  about 3 times shorter half-life (see Table \ref{tab:bbhalf}).

\begin{figure}[!ht]
\begin{center}
\includegraphics[width=0.6\textwidth,height=0.47\textheight]{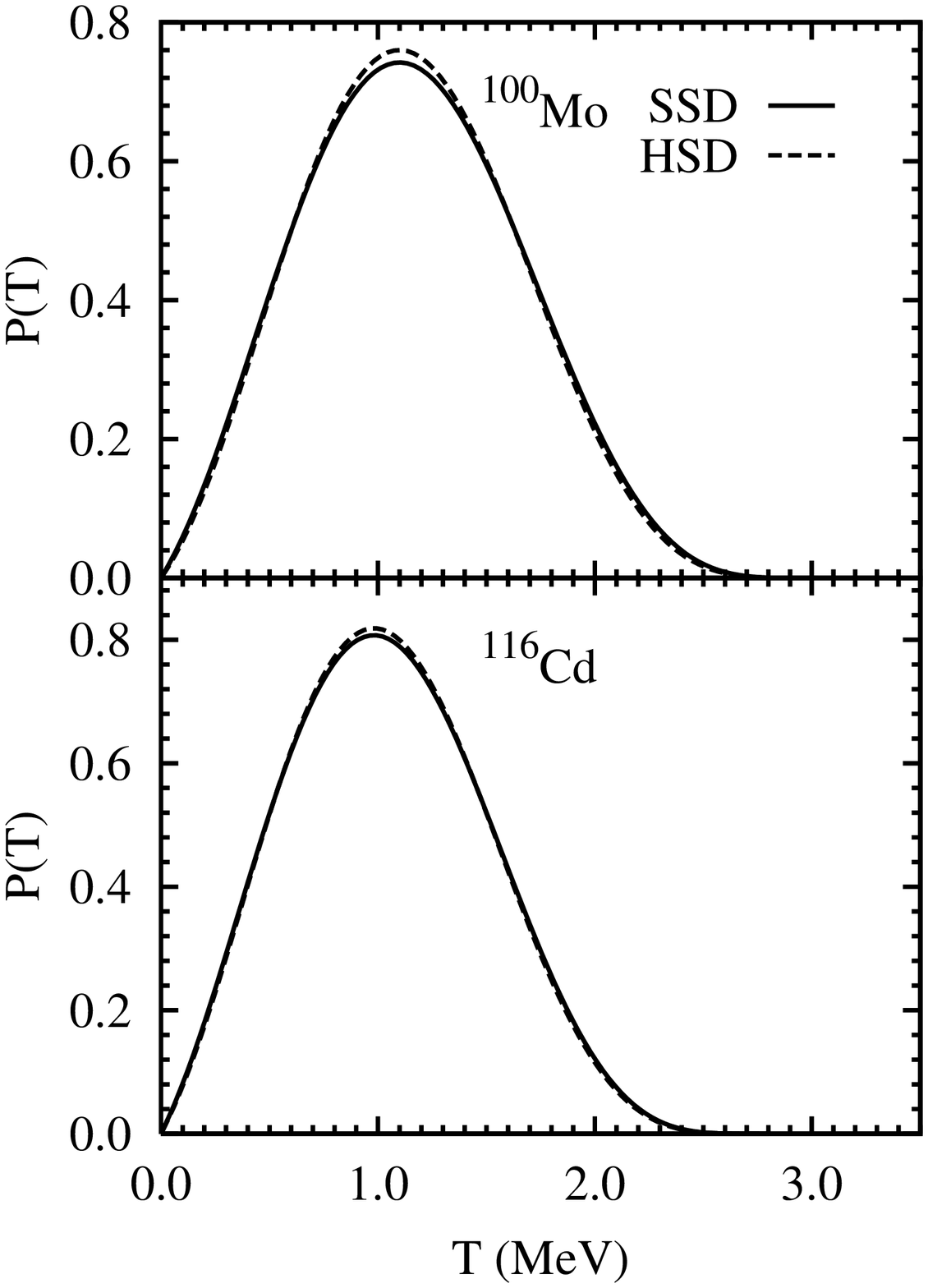}
\caption{
The differential decay rates normalized to the total decay rate 
$P(T) = (1/\omega) d \omega/d T$ vs. 
the sum of the kinetic energy of outgoing electrons 
$T = m_e (\varepsilon_1 + \varepsilon_2 - 2)$ for $2\nu\beta\beta$-decay of
$^{100}Mo$ and $^{116}Cd$ to the ground state of final nucleus.  
The conventions are the same as in Fig. \protect\ref{fig:gs2gs}.
\label{fig:sum}}
\end{center}
\end{figure}

The difference between the SSD and HSD predictions at the 
beginning of the single electron spectra is of the order of
one percent. 
Thus the experimental observation of this effect requires 
large experimental statistics.
The corresponding conditions have been reached in
the NEMO 3   experiment with  $^{100}Mo$
collected more than one hundred events of $2\nu\beta\beta$-decay 
of $^{100}Mo$  \cite{mo100g}. 
The obtained results are in favor of the SSD hypothesis for transition to the
ground state. There is an additional possibility for a cross-check
of this conclusion by analyzing the differential decay rate 
$P(T)$ [see Eqs. (\ref{eq:eneprob}), (\ref{2nbbT:S}) and (\ref{2nbbT:H})],
where $T$ is the sum of the kinetic energies of outgoing
electrons. For the $2\nu\beta\beta$-decay of $^{100}Mo$ and
$^{116}Cd$ the summed energy distributions are shown in Fig. 
\ref{fig:sum}. We see that the shape of HSD and SSD curves
is again quite different. The maximum of the HSD distribution
is slightly above the maximum of the SSD one. 
The large statistics to be collected by the NEMO 3 experiment 
in the near future  will allow us to study this effect as well.

\begin{figure}
  \begin{center}
\includegraphics[width=0.6\textwidth,height=0.7\textheight]{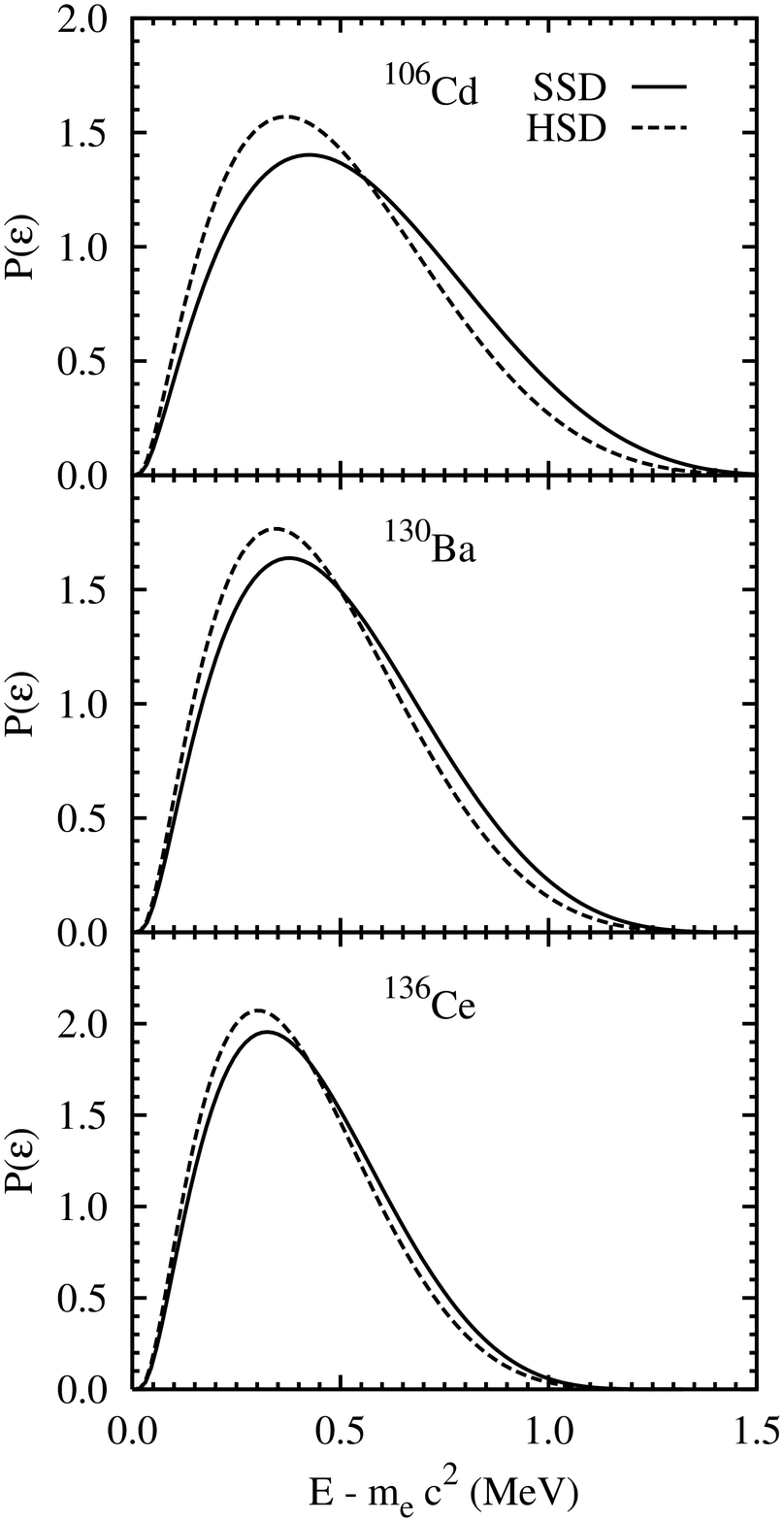}
\caption{The single positron differential decay rate 
normalized to the total decay rate 
$P(\varepsilon) = (1/\omega) d \omega/d \varepsilon $ vs. 
the positron kinetic energy $\varepsilon$ for
the ground state to ground state $2\nu\varepsilon\beta^{+}$-decay 
transitions 
${^{106}\mathrm{Cd}}\rightarrow {^{106}\mathrm{Pd}}$, 
${^{130}\mathrm{Ba}}\rightarrow {^{130}\mathrm{Xe}}$, 
and 
${^{136}\mathrm{Ce}}\rightarrow {^{136}\mathrm{Ba}}$.
The electron capture from the $K$ shell is assumed.
The conventions are the same as in Fig. \protect\ref{fig:gs2gs}.
 \label{fig:gs2gsec}}
\end{center}
\end{figure}

In Fig.~\ref{fig:gs2gsec} we show the single positron differential 
decay rate $P(\varepsilon)$ normalized to the total decay rate 
[see Eqs. (\ref{eq:eneprob}), (\ref{2nbbT:S}) and (\ref{2nbbT:H})]
for ground state to ground state $2\nu\varepsilon\beta^{+}$-decay 
transitions 
${^{106}\mathrm{Cd}}\rightarrow {^{106}\mathrm{Pd}}$, 
${^{130}\mathrm{Ba}}\rightarrow {^{130}\mathrm{Xe}}$, 
and 
${^{136}\mathrm{Ce}}\rightarrow {^{136}\mathrm{Ba}}$.
One can notice
a significantly larger difference between
SSD and HSD distributions in comparison with the case of the
$2\nu\beta^{-}\beta^{-}$-decay of $^{100}Mo$
(see Fig. \ref{fig:gs2gs}). 
%
This is partially due to the fact that
in these two cases with emitted positron and emitted electrons
the behavior of the relativistic Coulomb factor 
as a function of energy is quite different.
In the case of positron its 
value is vanishing for kinetic energy equal to zero 
while for the electron it takes a maximal value. On the other hand
the predicted SSD half-lives 
of the studied $2\nu\varepsilon\beta^{+}$-decays
(see Table  \ref{tab:eechalf}) are about 2-3 orders of magnitude
larger than the measured half-life of
$2\nu\beta^{-}\beta^{-}$-decay of $^{100}Mo$. Thus, the experimental
study of the  $2\nu\varepsilon\beta^{+}$-decays with the purpose of 
verification of the SSD hypothesis looks unfavorable.
%

\section{\label{sec:sum}Summary and Conclusions}

In this paper we have systematically studied the SSD hypothesis for all
$\beta\beta$-unstable isotopes with the $1^+$ spin-parity ground state of
intermediate nucleus presented in Tables \ref{tab:bbdata}, \ref{tab:eedata}. 
The two-neutrino accompanied $\beta^-\beta^-$, $\beta^+\beta^+$, 
$EC\beta^+$ and $EC/EC$ modes
of double beta decay to ground ($0^+$) and excited ($0^+$, $2^+$) states of 
final nucleus were considered.
The half-lives of these processes were derived within the
SSD hypothesis using the values of the matrix elements of 
single-$\beta^-$ and electron capture decays extracted from the measured 
$log ft$ values. In those cases when only the lower limits 
on the $log ft$ value were available, we calculated the lower 
limits on the corresponding double beta decay half-lives. 
The SSD calculations of $\beta\beta$-decay rates
were performed without additional approximations 
keeping explicitly the dependence of energy denominators on the lepton 
energies \cite{sim01}. 
In comparison with the previous studies~\cite{sem00,sim01}
we have more accurately taken into account the effect of Coulomb suppression
in the part which corresponds to the charge distribution of final nucleus
and described by the correction factor $R_{11}^{\pm}(\varepsilon)$ 
(see Eqs. \ref{ecoulf}, \ref{eq:26}). 
We have found that the improved treatment of the Coulomb interaction of
emitted charged lepton with the nucleus modifies the total rate of 
$2\nu\beta^{-}\beta^{-}$-decay in $5 - 10\%$. 
For the case of the $2\nu\beta^{+}\beta^{+}$-decay this improvement results 
in  $20 - 50\%$ enhancement of the half-life(Tab.~\ref{tab:bbthalfr11})

We demonstrated that the existing experimental data on the half-lives 
of the $2\nu\beta^{-}\beta^{-}$-decay 
of  $^{100}Mo$ (to $0^+$ ground \cite{mo100g} and $0^+_1$ excited
\cite{mo100e} states) and of $^{116}Cd$ (to $0^+$ ground state 
\cite{cd116}) show a clear
tendency to favor the SSD hypothesis.
However, we are not yet in the position to make definite conclusion 
on the validity of the SSD hypothesis. This is, partially, because
of quite large uncertainties (about $50\%$) in the SSD predictions 
stemmed from insufficient precision of the existing experimental measurements
of the $log ft_{EC}$ values for electron capture.
Thus, a more accurate experimental information on the
associated $\beta$- and $\beta\beta$-transitions is needed.

The SSD approach allowed us to make the predictions for those
neutrino (antineutrino) accompanied $\beta\beta$ decays
to ground and excited final states, which have not yet been
experimentally observed.
Despite the fact that these predictions should be taken as 
order-of-magnitude estimates
they may help in evaluation of the prospects of experimental searches for 
mentioned processes.

The SSD results presented in Tables \ref{tab:bbhalf}-\ref{tab:ecechalf} 
show that there is a chance to observe $2\nu \varepsilon\varepsilon$-decay
for $^{106}Cd$, $^{112}Sn$ and $^{130}Ba$ at the level of
$10^{21}-10^{22}$ years. 
On the other hand for $2\nu\varepsilon\beta^{+}$- and 
$2\nu\beta^{+}\beta^{+}$-decays the SSD predicts, in general, 
significantly larger  
half-lives looking quite pessimistic from the view point of the 
experimental observation of 
these processes in the near future.
We also found that with the exception of the
$2\nu\beta^{-}\beta^{-}$-decay of $^{100}Mo$
to $0^+_1$ excited state the 
SSD half-lives of all the studied
transitions to excited states $0^+$ and $1^+$ states 
are above $10^{23}$ years. 
 
We proposed the study of the differential decay rates as a new perspective 
possibility for the experimental verification of 
the SSD hypothesis. 
One theoretically important point with respect to these 
characteristics consists in the fact that the differential  
rates normalized to the corresponding full decay rates 
do not depend on the values of the associated nuclear matrix elements.
We have shown that the best candidates for this study 
are double beta decay chains with low-lying $1^+$ ground states of
intermediate nucleus. 
In this case the shape of the single electron/positron distribution
and the summed electron spectrum calculated within the SSD 
are very sensitive to the lepton energies present in the energy denominators.
On the other hand, if the main contribution to the double beta decay
matrix elements comes, contrary to the SSD hypothesis, 
from the transition through the higher lying states of intermediate nucleus
(HSD hypothesis) the dependence
on lepton energies in the energy denominators is negligible resulting
in the distributions which can be experimentally distinguished from 
the SSD case. In the present paper we have shown that the precision  
measurements of the differential characteristics 
of the two-neutrino modes of the
$\beta^-\beta^-$-decay of $^{100}Mo$ and 
$^{116}Cd$ and $EC/\beta^+$-decay of  
$^{106}Cd$, $^{130}Ba$ and $^{136}Ce$
are able to confirm or rule out the SSD hypothesis.
Recently the NEMO 3 collaboration have started the data analysis
aimed to distinguish the SSD and HSD predictions for 
the $2\nu\beta^-\beta^-$-decay of $^{100}Mo$ \cite{mo100g}. 

We believe that this and further studies of the SSD hypothesis will
allow us to improve the understanding of the structure of 
two-neutrino  double beta decay matrix elements 
and will be helpful in fixing the nuclear 
structure parameter space needed for the 
calculations of the matrix elements 
of neutrinoless double beta decay
within the most advanced nuclear models \cite{rodin}.

\begin{acknowledgments}
This work was supported by the FONDECYT project 1030244,  
by the President grant of Russia 1743 ``Scientific Schools", 
by the VEGA Grant agency of the Slovac Republic under the contract 
No.~1/0249/03, by the EU ILIAS project under the contract
RII3-CT-2004-506222 and by the Grant Agency of the Czech Republic Grant No.
202/02/0157.
\end{acknowledgments}

\end{document}